\documentclass[aip,jcp,numerical,
preprint,
amsfonts,floatfix]{revtex4-1}
\usepackage{amsmath}    
\usepackage{graphicx}   
\usepackage{verbatim}   
\usepackage{color}      
\usepackage{subfigure}  

\begin{document} 

\title{Analysis of the Equilibrium and Kinetics of the Ankyrin Repeat Protein Myotrophin}

\author{Mauro Faccin}

\author{Pierpaolo Bruscolini}
\affiliation{Departamento de F\'{\i}sica Te\'{o}rica \& Instituto de Biocomputaci\'{o}n y F\'{\i}sica de Sistemas Complejos (BIFI),
Universidad de Zaragoza, Edificio I + D, c/ Mariano Esquillor s/n - 50018 Zaragoza
(Spain)}
\email{pier@unizar.es}

\author{Alessandro Pelizzola}
\affiliation{Dipartimento di Fisica, CNISM Unit\`a di Torino and INFN,  Politecnico di Torino, Corso Duca degli Abruzzi 24, 10129
Torino (Italy)}


\begin{abstract}
We apply the Wako-Saito-Mu{\~n}oz-Eaton model to the study  of Myotrophin, a small ankyrin repeat protein, whose folding equilibrium and kinetics  have been recently characterized experimentally. The model, which is a native-centric with binary variables, provides a finer microscopic detail than the Ising model, that has been recently applied to some different repeat proteins, while being still amenable for an exact solution. 
In partial agreement with the experiments,  our results reveal a weakly three-state equilibrium and a two-state-like kinetics of the wild type protein despite the presence of a non-trivial free-energy profile. These features appears to be related to a careful ``design'' of the free-energy landscape, so that mutations can alter this picture, stabilizing some intermediates and changing the position of the rate-limiting step. 
Also the experimental findings of two alternative pathways, an N-terminal and a C-terminal one, are qualitatively confirmed, even if the variations in the rates upon the experimental mutations cannot be quantitatively reproduced. 
Interestingly, folding and unfolding pathway appear to be different, even if closely related: a property that is not generally considered in the phenomenological interpretation of the experimental data.
%
%
\end{abstract}

\date{\today}

\maketitle

\section{INTRODUCTION}
In the last decade, repeat proteins have increasingly drawn the attention of researchers, due to their ubiquity, their abundance, and the fact that they provide a different folding paradigm with respect to the well known one of globular proteins, where complex native state geometries, characterized by local and nonlocal interactions, are most often associated to a simple two-state equilibrium and kinetics.
On the contrary, repeat proteins are characterized by tandem arrays of the same structural motif (even if individual repeats show just partial sequence identity, typically, around 25\% \cite{Kloss2008}). Such motifs are usually arranged in a linear fashion, giving rise to elongated structures that may consist of a highly variable number of repeats. Interactions in such modular structures take place within a repeat and between adjacent repeats, while truly non-local interactions connecting non-contiguous repeats are lacking. While such organization provide a general-purpose scaffold that can be tuned to bind different species, it is quite suprising that it is still compatible with a cooperative, two-state folding. Indeed, recent experimental studies have revealed that repeat proteins typically show a two-state equilibrium but a multistate kinetics \cite{Kloss2008}, driving the attention on the existence of different folding pathways. From a theoretical point of view, repeat proteins provide an ideal framework for modeling and hypothesis-testing, due to their structural modularity, and to the fact that artificial molecules can be built from consensus sequences, so that the role of the different interactions and of the chain length can be dissected and analyzed individually.
Not surprisingly, the classical  Ising model from statistical mechanics has been used to  describe these almost linear systems with local nearest-neighbour interactions, where the spin variables have been identified with individual helices within a repeat \cite{Kaj2005}, or with entire repeats \cite{Wetzel2008}, or with the elementary ``foldons'' identified in a more detailed molecular dynamics simulation \cite{Ferreiro2008}. Typically, the external fields and neighbour interaction parameters ($h_i$ and $J_{i,i+1}$ respectively, in their typical textbook denominations) are derived from the experimental analysis of the stability of constructs
of different length, and are related to the variation of the areas accessible to the solvent in the folding process.
The identification of the elementary spin variable with a piece of structure as a whole, hinders the possibility to investigate the detailed role of individual contacts between the residues, and of studying the origin of the cooperativity and multistate kinetics on the residue scale. 

Here, we use the Wako-Saito-Mu{\~n}oz-Eaton (WSME) model \cite{WAKO1978,WAKO1978a,Munoz1998,Munoz1997,Munoz1999}, where the state of each residue $i$ is described by a binary variable $m_i=0,1$, representing the unfolded and native state, respectively. Formally, the model differs from the Ising one in that the interactions are not limited to next neighbours, but extend to any distance, provided that the variables corresponding to all the intervening residues are set to the native state. The model equilibrium can be exactly calculated \cite{WAKO1978,WAKO1978a,Bruscolini2002,Pelizzola2005}, so that energies, free-energies, and fractions of native residues can be easily evaluated. The folding and unfolding kinetics are studied through Monte Carlo simulation, with an elementary step corresponding to the folding/unfolding of one residue. 

The model has been applied to describe the folding of many proteins \cite{Zamparo2009,Cellmer2008,Itoh2008,Itoh2010,Abe2009,Morozov2007,Chung2008,Bruscolini2007a,Bruscolini2007},  and also to the study of force-induced denaturation of proteins and RNA \cite{Imparato2007,Imparato2007a,Imparato2008,Imparato2009,Caraglio2010}. We apply the WSME model to the study of Myotrophin, a 118 residues protein, ubiquitously expressed in all mammalian tissues \cite{Sen1990,Taoka1994,Sivasubramanian1996,Anderson1999}, made up of four ankyrin repeats. Its equilibrium has been characterized experimentally as two-state  by Peng and coworkers\cite{Mosavi2002a}  
with thermal and chemical denaturation experiments, and later confirmed as such, at least as far as chemical denaturations is concerned, by  Lowe and Itzhaki \cite{Lowe2007a}, that also studied the kinetics \cite{Lowe2007a,Lowe2007}. In the former paper, the authors propose an effective two-state framework to interpret the relaxation kinetics \cite{Lowe2007a} (more precisely, they actually observe some curvature in the unfolding arm of the chevron plot, that can be explained by postulating either a barrier shift or the existence of a high energy intermediate of negligible population).

 An extended analysis on several mutants leads them to conclude that, in order to explain within a unique framework the behavior of both the wild type and the mutants, pathway heterogeneity must be assumed, with the dominant pathway presenting a high energy intermediate, which is lacking in the secondary one \cite{Lowe2007}. 
Even if their analysis contains several simplifying assumptions (for instance, the fact that the relaxation rate is just the sum of the rates along the two pathways) they are able to provide very good fits to the experimental data, and to determine that the two pathways present different nucleation sites, on the N-terminal or on the C-terminal part of the protein, respectively. Finally, they show how, by combining mutations, it is possible to make the protein switch between the two pathways.

After fitting the model parameters  to reproduce the fraction of native protein as a function of the denaturant concentration derived from experiments, we calculate  the free energy profiles and relaxation rates, and characterize the relaxation pathways, at low and high denaturant concentrations, for the wild type protein and for a series of mutants, selected to probe different regions and contact distances. We also simulate the set of mutations used in Ref.~\cite{Lowe2007}, to test the double pathway hypothesis.

Our goal is to  reproduce, at least qualitatively, the experimental behavior, and to shed light on the nature of the folding nuclei, as well as to recover the role of ``pathway switch'' played by some mutations. Moreover, we want to clarify the different role played by mutations affecting local or non-local contacts in the same region.

\section{METHODS}
\label{sec:methods}
\subsection*{Model}

WSME is a native-centric model \cite{Ueda1978}, i.e. it relies on the knowledge of
the native state of a protein to describe its equilibrium and kinetics. 
Its binary variables $m_k$, accounting for the local backbone and side chain angles, describe the state of each residue  $k \in [1, N] $ 
 as ordered (native,  $m_k=1$) and disordered
(unfolded, $m_k=0$). Since the latter state allows a much larger number of
microscopic realizations than the former, an entropic cost  $q_k$ is
given to the ordering of residue $k$.

\begin{figure}[!ht]
\includegraphics[width=0.3\textwidth,angle=-90]{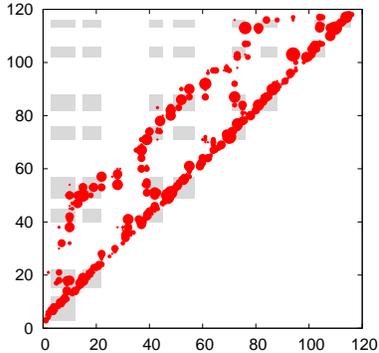}
\caption{\label{fig:map}Weighted contact map for myotrophin (2myo). The
area of each circle is proportional to the weight of the contact between
residues $i$ and $j$, in terms of the number of interatomic interactions (see text for details). Darker areas represent contacts between residues that
belong to helices.}
\end{figure}

The model is described by the 
effective hamiltonian (indeed, a free energy, where the solvent and the fast degrees of freedom have been integrated out):
\begin{equation}
H =  - \sum_{i=1}^{N-1}\sum_{j=i+1}^{N}
\epsilon_{i,j} \Delta_{i,j} \prod_{k=i}^{j} m_k
+  \sum_{k=1}^{N} (q_k  T +  \alpha c) m_k  ,
\label{Hamiltonian}
\end{equation}
where $N$ is the number of residues  in the
molecule 
and $T$ the absolute
temperature. The
product $
{\prod_{k=i}^j} m_k$ takes value 1 if and only
if all the peptide bonds from $i$ to $j$ are in the native state,
thereby realizing the assumed interaction. Non--native interactions are disregarded, while native interactions are accounted for in the contact matrix $\Delta_{i,j}$, which counts the number of contacts between atoms of non--contiguous residues $i$ and $j$ in the native structure, according to a cutoff distance criterion. In the following, we will use the contact map calculated from the crystal structure of Myotrophin deposited in the Protein Data Bank  (PDB code: 2myo), considering that a contact is established if any two atoms (including hydrogens) from residues $i$ and $j$ are found at a distance less than 3.5 $\mathring{A}$.  Figure \ref{fig:map} reports the resulting contact map.

The expression above differs from the original one for the  last term, accounting for the interaction between the denaturant and the protein backbone (as suggested by Bolen and coworkers\cite{Auton2007}, and also in agreement with the choice in Ref.~\cite{Ferreiro2008}), where $c$ represents the  urea molar concentration and $\alpha$ is a new parameter. 
For the sake of simplicity,  we take homogeneous parameters $\epsilon_{i,j} =  \epsilon$, $q_i=q$, for each $i$ and $j$, with $\epsilon,q>0$, to model the wild type protein. 
We use the experimental thermodynamics data to adjust the parameters: we set $q=2 R$ 
and find $ \epsilon$, and $\alpha$ by fitting the equilibrium experimental data of Refs.~\cite{Mosavi2002a,Lowe2007a} for the wild type species. We first calculate the native and unfolded baselines $n(z)$ and $u(z)$, where $z$ is the temperature or denaturant concentration, from a linear interpolation of the data far from the transition region. Then, we consider the order parameter:
\begin{equation}
 p(z)=\frac{m(z)-u(z)}{n(z)-u(z)}\,,
\label{eq:ordpar_p}
\end{equation}
normalized between zero (unfolded) and one (native). Here $m(z)$ is the equilibrium average fraction of folded residues, defined  in Eq.~(\ref{eq:avenatfrac}).
Then, we adjust the $\epsilon$ parameter, imposing that the temperature $T_m$ at which $p(T_m)=0.5$ coincides with the experimental mid-folding temperature $T_m=327$~K . Then, we do the same for the $\alpha$ parameter, imposing that $p(c_m)=0.5$ at the experimental mid-folding denaturant molar concentration $c_m=3.2$. The resulting values of the parameters are used in the whole study, for both wild type and mutated species.

\begin{figure}[!ht]
\centering
\subfigure[{\protect [Urea]}]{
\includegraphics[width=0.26\textwidth,angle=-90]{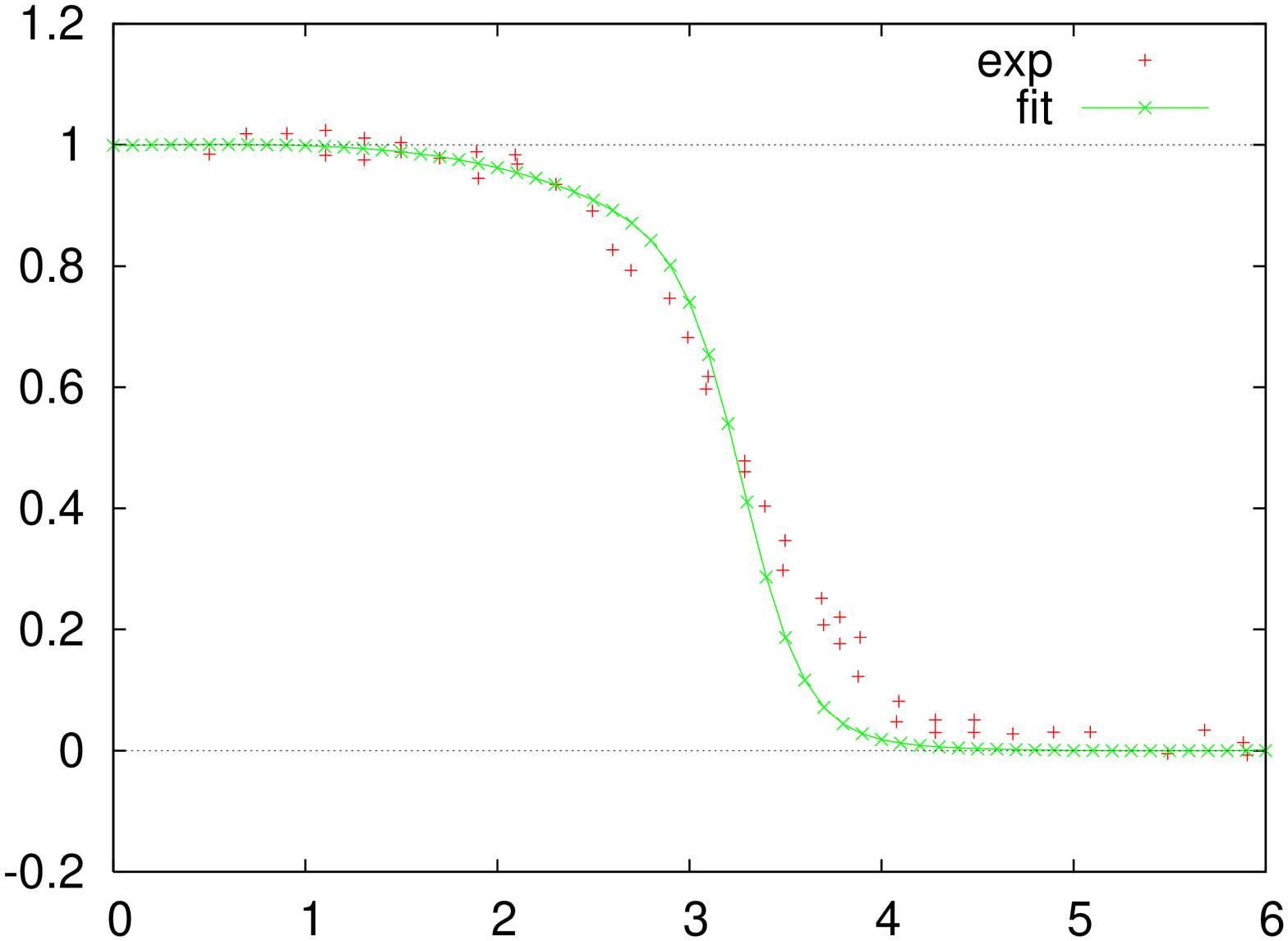}}\\
\subfigure[Temperature]{
\includegraphics[width=0.26\textwidth,angle=-90]{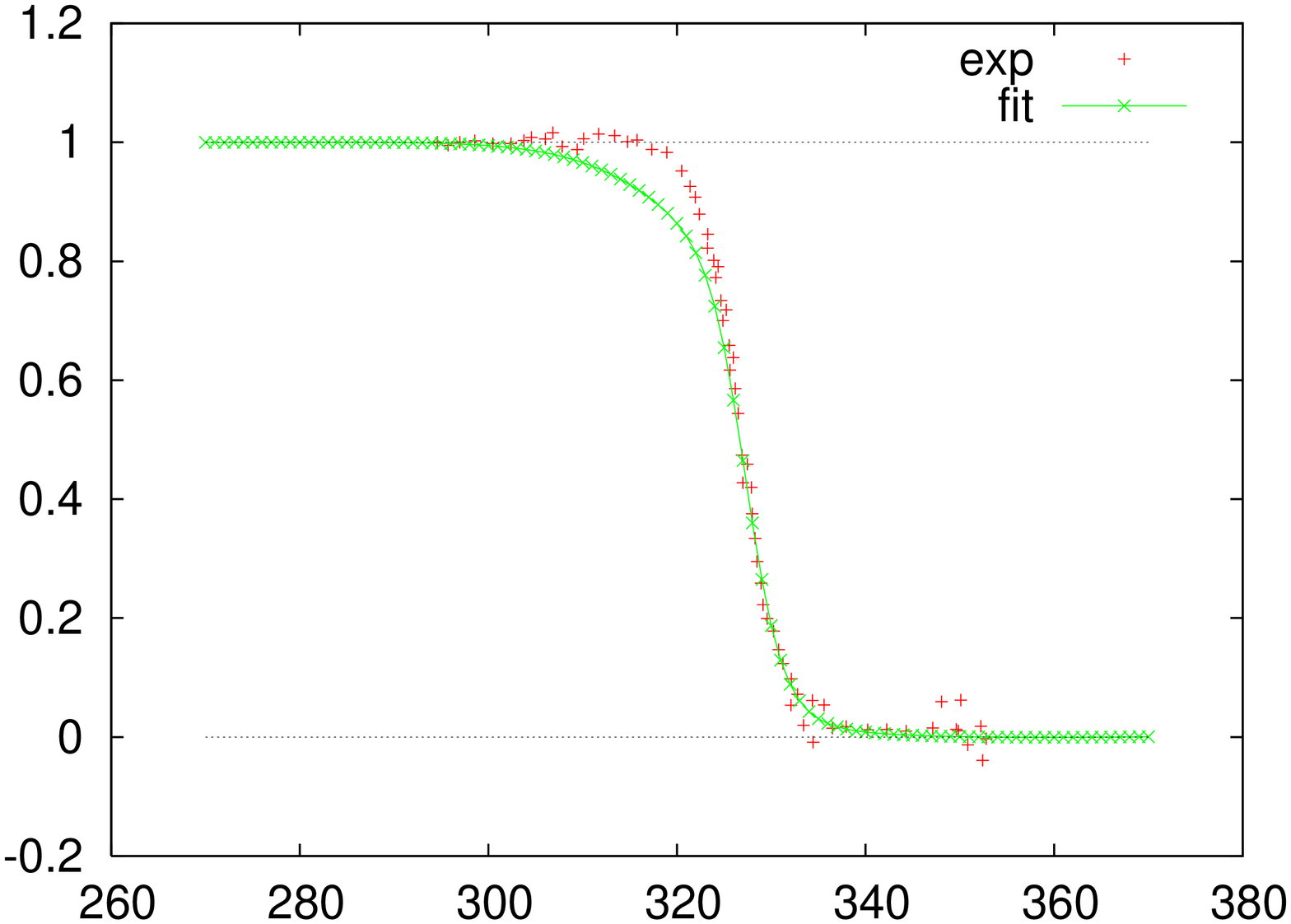}}\\
\caption{
\label{fig:den-fit} 
Fit of the order parameter $p(z)$ from the WSME model to the experimental data for (a) urea-induced denaturation{\protect \cite{Lowe2007a}} and (b) thermal denaturation{\protect \cite{Mosavi2002a}}. 
Parameters values are: $\varepsilon=0.2276$ kJ/mol, 
$q=0.0166$  kJ/(K mol), $\alpha=0.208$ kJ/([urea] mol)}
\end{figure}

The results are reported in Fig.~\ref{fig:den-fit}.

Mutations are mimicked by perturbing a group of contacts, of one or more residues as detailed below, through the addition of a $\Delta \epsilon_{i,j}$ to the corresponding interactions.   To make comparison easier, the same total perturbation of $\varepsilon_{tot}$=9.21 kJ/mol, comparable with those reported in Ref.~\cite{Lowe2007a},  is introduced for all mutants, so that the $\Delta \epsilon_{i,j}$ will vary between  mutants, according to the number of affected contacts $n_{ct}$: $\Delta \epsilon_{i,j}=\varepsilon_{tot}/n_{ct}$. 
The list of analyzed mutations is reported in Table~\ref{table:mutations}, and was selected to probe different regions of residues and different distances between contacting residues. 
\begin{table}[!htb]
\begin{tabular}{ c | l }
\hline 
name & description \\
\hline
WT		& wild type protein\\
$S_{1,2}$	& contacts of residues [5\dots10] with residues [17,18]\\
$S_{3}$		& contacts of residue 32\\
$S_{3,4}$	& contacts of residues [36\dots44] with residues [49\dots56]\\
$S_{1,4}$	& contacts of residues [9\dots18] with residues [45\dots53]\\
$S_{5,6}$	& contacts of residues [71\dots76] with residues [82\dots88]\\
$S_{3,6}$	& contacts of residues [42\dots52] with residues [78\dots83]\\
$S_{7}$		& contacts of residue [103] with residues [94\dots101]\\
$S_{7,8}$	& contacts of residues [104,105] with residues [113,114]\\
$S_{5,8}$	& contact  between residue 76 and residue 113\\
\hline \hline
\end{tabular}
\caption{\label{table:mutations}  List of mutations analyzed in this work. The same total destabilization of 9.21 kJ/mol was used in all cases, adapting the individual $\Delta \epsilon_{i,j}$ of each contact. The indices $m,n$ in $S_{m,n}$ specify the helices involved in the mutated contacts. $S_3$ actually affects a loop residue, close to helix 3.}
\end{table}

We also analyze the case of some multiple mutations that have been investigated experimentally in Ref.~\cite{Lowe2007} to test the two-pathway interpretation.  In order to compare our model to those results, we have simulated the effect of those mutations applying a stabilizing or destabilizing perturbation (whose energy is taken from Ref.~\cite{Lowe2007}) equally spread on
all the contacts of the mutated residues. The following
mutants are considered

\begin{itemize}
\item E17V/D20L ($\Delta\Delta G_\textrm{eq.}= -7.12 - 2.09$ kJ/mol C: a stabilized mutant)
\item A9G ($\Delta\Delta G_\textrm{eq.}= 10.38$ kJ/mol C)
\item A115G ($\Delta\Delta G_\textrm{eq.}= 3.39$ kJ/mol C)
\item A115G/A9G
\item A115G/E17V/D20L
\item A9G/E17V/D20L/A115G
\end{itemize}

Multiple mutations are considered as independent, and the corresponding energetic perturbation is applied to each point mutation separately, so that, e.g., $\Delta \Delta G_{A115G/A9G}= \Delta \Delta G_{A115G} + \Delta \Delta G_{A9G}$.

\subsection*{Thermodynamics}
The equilibrium values of all thermodynamic quantities  are calculated resorting to the exact solution of the model \cite{Bruscolini2002,Pelizzola2005}. In particular, we will study the fraction of native residues:
\begin{equation}
m=\frac{1}{N} \sum_{i=1}^{N} \langle m_i \rangle\,,
\label{eq:avenatfrac}
\end{equation}
and
the free-energy profiles as a function of the number of native residues $M$:
\begin{equation}
F(M)=-RT \log{Z_M}\,,
\label{eq:fprofile} 
\end{equation}
where $Z_M=\sum'_{\{ m_i\}} \exp{(- H/RT)}$, and the sum $\sum' (\bullet)$ is restricted to the states with a fixed number of native residues $\sum_i m_i=M$. The $Z_M$ can be easily calculated within the framework of the exact solution mentioned above. The reaction coordinate is defined as $\rho=M/N$.  
Finally, we will study the average values  
\begin{equation}
\mu_{i,j} = \langle m_{i,j} \rangle \,,
\label{eq:mu-strings}
\end{equation}
and
\begin{equation}
 \nu_{i,j} = \langle  (1-m_{i-1}) m_{i,j}  (1-m_{j+1}) \rangle\,,
\label{eq:nu-cappedstrings}
\end{equation}
of  the products $m_{i,j}=\prod_{k=i}^j m_k$. Since these products take on value 0 or 1, $\mu_{i,j}$ and $\nu_{i,j}$ represent the equilibrium probability that the region between $i$ and $j$ is native (and, in the second case, that it is capped by unfolded residues, thus representing an isolated native region).


\subsection*{Kinetics}
The kinetic evolution of the model  is described through a
discrete--time master equation, $p_{t+1}(x) = \sum_{x'} W(x' \to x) p_t(x)$, for the probability distribution
$p_t(x)$ at time $t$, where $x = \{ m_k, \,k = 1, \ldots N \}$ denotes
the state of the system. 
Unfortunately this expression is not amenable to analytical treatment (even if an accurate semi-analytical
approximation exists \cite{Zamparo2006,Zamparo2006a}), since by construction $W(x' \to x)$ is a $2^N \times 2^N$ matrix.  
Here the kinetics will be studied by means of Monte Carlo simulations: as in previous works \cite{Zamparo2006,Zamparo2006a}, the transition matrix $W$ is specified by a single bond flip Metropolis rule, which implies that a flip is accepted or rejected according to its equilibrium probability, at the temperature and denaturant concentration specified for the simulation. 
We study the kinetics in both folding (T=293.15 K, $c$=0) and unfolding conditions (T=293.15 K, $c$=12): in folding simulations, the initial state is a random configuration extracted with the infinite temperature equilibrium probability, so that the initial
and final fraction of native residues are $ m(t=0)=0.12$ 
and 
$m(t=\infty)=0.97$ respectively, for the wild type protein (slightly different values are obtained for the mutants). 
In unfolding simulations, the fully native state ($m_i=1$ for each $i$) is assumed as the initial condition, while $m(t=\infty)=0.0047$, 
for the wild type protein.


We study the relaxation of the average fraction of native residues $m(t)$: at each time, the average is formally calculated as in Eq.~(\ref{eq:avenatfrac}), but with the $\langle \bullet \rangle$ now indicating the average over  $\cal N$ single molecule simulations, that is, over an ensemble  of $\cal N$ molecules. We choose  $\cal N$=2000 as a reasonable tradeoff between detecting a neat signal and reducing simulation time.  We fit $m(t)$  with one- or two-exponential expressions, namely:
\begin{equation}
m(t)=m_{eq}(T,c) + c_1 e^{-k_1 t} \,,
\label{eq:1expfit} 
\end{equation}
or
\begin{equation}
m(t)=m_{eq}(T,c) + c_1 e^{-k_1 t} + c_2 e^{-k_2 t}\,,
\label{eq:2expfit} 
\end{equation}
where $m_{eq}(T,c)$ is the equilibrium value at the temperature  $T$ and denaturant concentration $c$, obtained from the thermodynamics calculations. The fitting parameters are the rates $k_i$ and the corresponding amplitude $c_i$. 

We also characterize the folding pathways, by observing  the average times of structure formation: we consider the regions $h_l=(i_l,j_l)$, $l=1,\dots,8$ corresponding to the eight helices of native myotrophin, as well as the regions $R_{\alpha,\beta}$ encompassing the fragment from helix $\alpha$ to $\beta$ inclusive (that is, from residue $i_{\alpha}$ where helix $\alpha$ begins, to residue $j_{\beta}$ where helix $\beta$ ends). 
After defining the folding time $t_f$ as the first passage time through the state with all the helices formed ($R_{1,8}$), we identify, for each single molecule simulation of the folding process, the stabilization time  $t_{\alpha,\beta}^{(f)}$ of each region $R_{\alpha,\beta}$ as the last time it turns completely native (thus, waiting for all the fluctuations to fade away). This choice is a natural generalization of that proposed in Ref.~\cite{Zamparo2009}, to the present case with many elements of secondary structure: notice indeed  that, due to the model characteristics, the stabilization of $R_{\alpha,\beta}$ in the native conformation is a necessary and sufficient condition for the formation of contacts between helix $\alpha$ and $\beta$ (if any), as well as between all pairs of helices $k$,$l$, with $\alpha \le k < l \le \beta$. The determination of $t_{\alpha,\beta}^{(f)}$ for all regions allows us to determine pathways in the secondary structure formation, and to identify two main pathways in the folding and unfolding of myotrophin (see Section \ref{sec:results} below).  

We also record the joint probabilities that two (non-overlapping) regions  are native at the same time, for each single-molecule simulation. This is to avoid that a wildly fluctuating element, with a late stabilization,  induce an artificial ordering along the pathway. We have observed, in any case, that the only elements in myotrophin for which strong  fluctuations could induce a problem are the first and last helix, which on the other hand turn out to be unimportant for pathway determination (see Section \ref{sec:results}). For the other helices, we observe that local fluctuations can indeed invert the order by which a region is stabilized,  in different single-molecule runs, starting from its constituent elements. However, the difference in stabilization times among the latter is small, allowing to group clearly which elements stabilize basically altogether in the folding process.  

We do the same for the unfolding simulations: now the unfolding time $t_u$ is defined as the first passage time in a state with $m<0.09$, and for each single molecule simulation, we record the last time $t_{\alpha,\beta}^{(u)}$ in which each region $R_{\alpha,\beta}$ switches from the native to unfolded state.

\section{RESULTS}
\label{sec:results}

\subsection{EQUILIBRIUM}
\subsubsection{Myotrophin presents a multi-minima free-energy profile, yet a sigmoidal equilibrium signal. 
}
\begin{figure}[!ht]
\centering
\subfigure[]{
\includegraphics[width=0.26\textwidth,angle=-90]{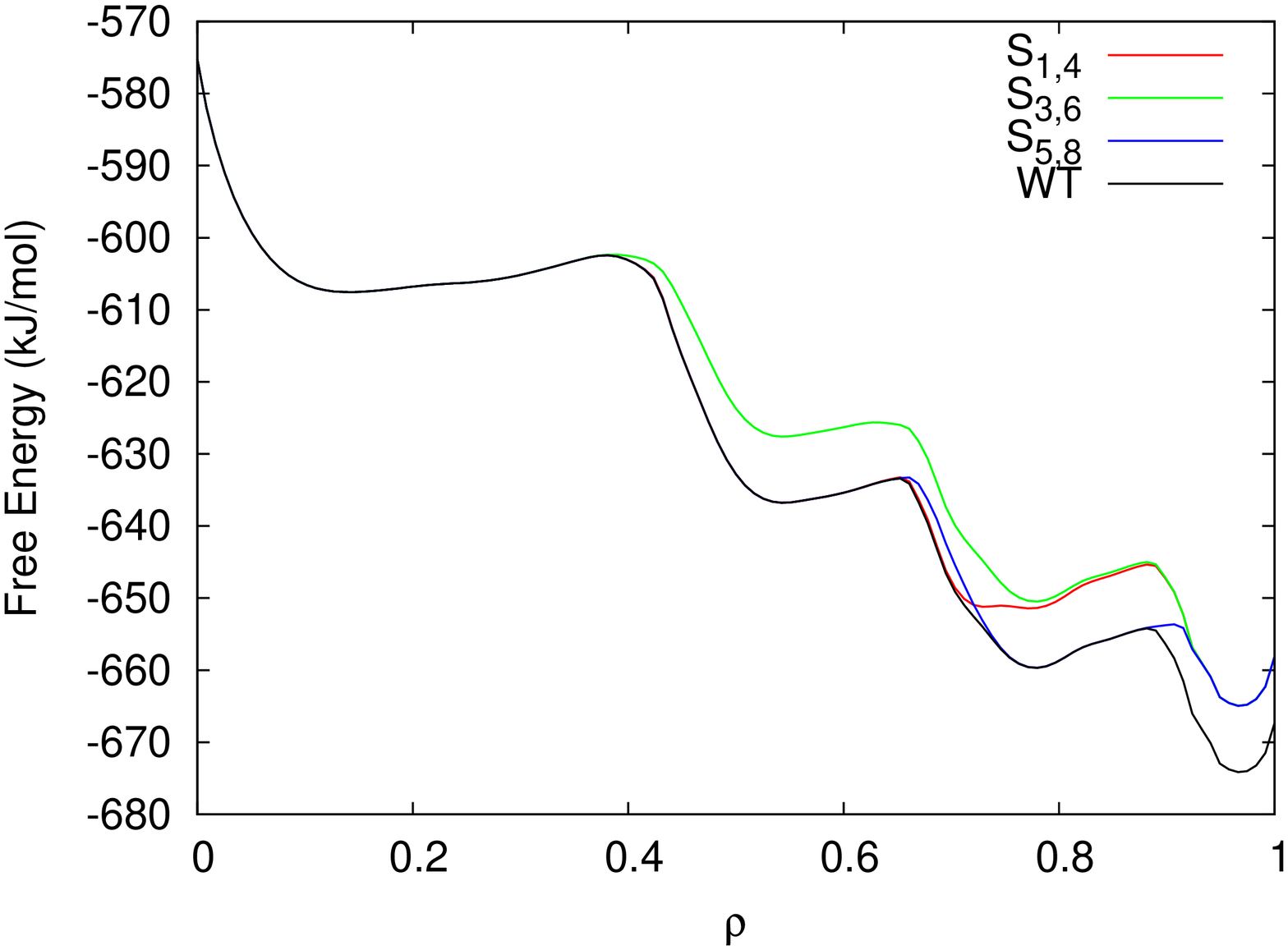}}\\
\subfigure[]{
\includegraphics[width=0.26\textwidth,angle=-90]{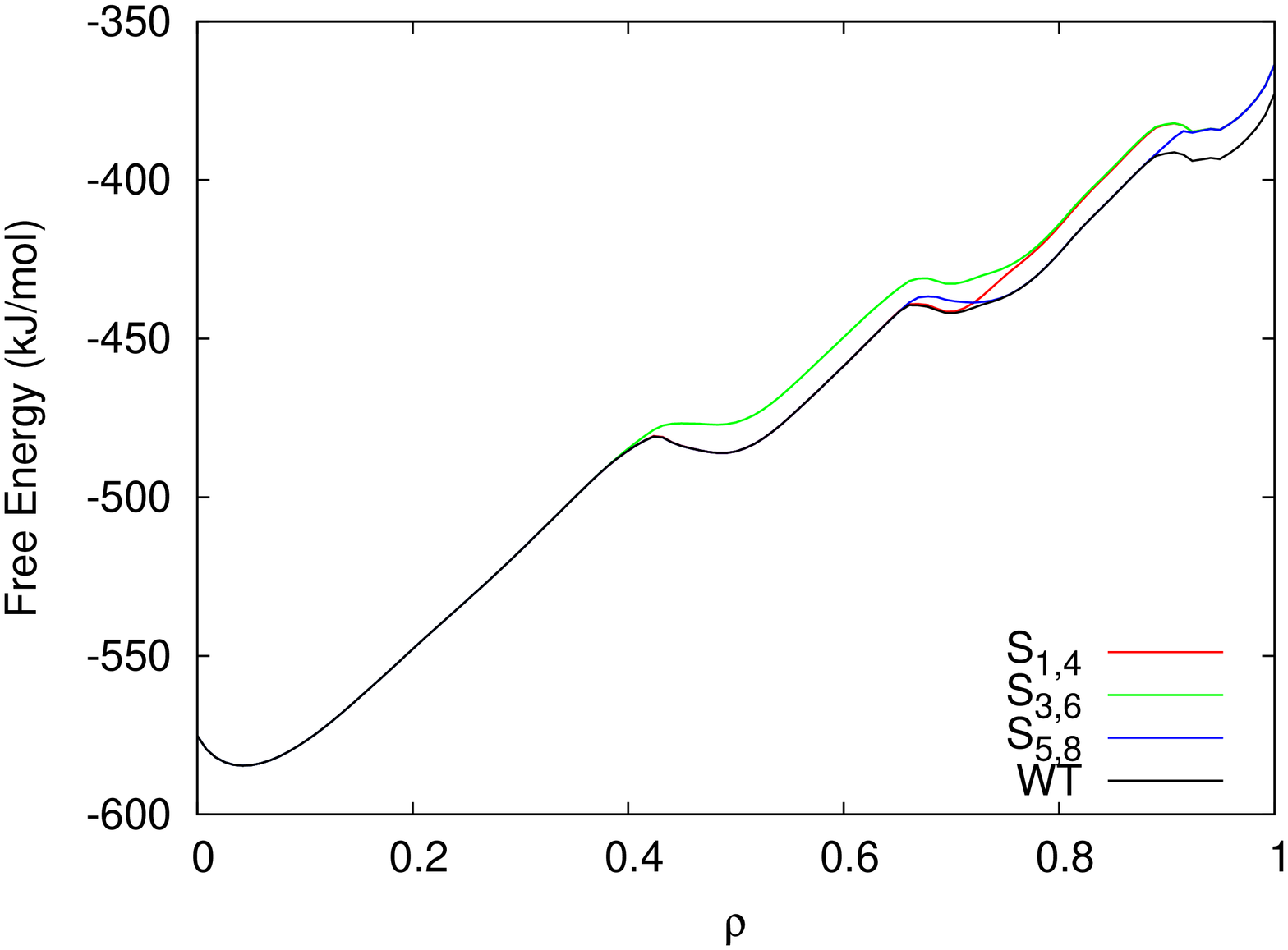}}\\
\caption{ Examples of the Myotrophin free-energy profiles as a function of the fraction of native residues $ \rho = M/N$. Panel a: renaturing conditions (`$c$=0). Panel b: denaturing conditions ($c$=12).}
\label{fig:eq}
\end{figure}

Figure \ref{fig:den-fit} reports the signals for the order parameter defined in Eq.~\ref{eq:ordpar_p}, as a function of denaturant concentration and temperature. 
Notice that such order parameter, at difference with some common experimental techniques, provides a global information on the protein behavior, and the sigmoidal shape of its signals (which are even sharper than the experimental data, probably due to the model energy function, that tends to enhance cooperativity \cite{Abe2006}) suggests 
a two-state interpretation. However, 
the analysis of the free-energy profiles as a function of the number of native residues $M$  reveals four minima, in both strongly renaturing and denaturating conditions (Fig.~\ref{fig:eq}). 

This apparent puzzle is solved by observing that, in the wild type species, the intermediate minima, for most temperature or denaturant concentration, are found at a free energy higher than the native or unfolded minima, and are not 
sufficiently populated to compete with them, resulting in an overall sigmoidal signal for the order parameter. 
Actually, the analysis of the profiles reveals that one intermediate becomes significantly populated in a small region close to the transition. Accordingly,  a more detailed inspection of the signal of the order parameter (see Ref.~\cite{suppmat1},  
Figs.~S2, S3) proves that a three-state fit yields an improvement in the accuracy of the fit which is statistically significant, according to the F-test.  
Mutations perturb the signal of the order parameter in position and/or in shape: 
in general, we have seen that destabilizing the central region lowers the mid-transition concentration, but basically preserves the degree of cooperativity implicit in the sigmoidal shape, while mutations at the N-term, and even more at the C-term, enhance the role of the intermediates, and can even induce a plateau; see Figure S4 in Ref.~\cite{suppmat1}. 

Figure \ref{fig:eq} reports also the behavior of three mutants: $S_{1,4}$, $S_{3,6}$, $S_{5,8}$, with mutation affecting the N-term, central region and C-term respectively.
It can be noticed that the free-energy profiles of the three mutants depart from the wild-type one at different values of the order parameter: while the unfolded minimum is unaffected by all mutations, indicating that none of the perturbed contacts is formed in the unfolded state, the destabilization of the central region affects weakly structured conformations as well, suggesting that this region contributes the most to the free-energy profile at low values of the reaction coordinate. Moreover, the profile at intermediate value of the reaction coordinate parallels that of the wild type: these configurations are evenly perturbed by the mutation. On the other hand, the mutation perturbing the C-term only affects the native minimum (suggesting that the C-terminal helix just gets stabilized in the native structure) and the  second barrier, with effects that will be clear in Section \ref{subsec:dynamics}. 
Finally, perturbing the N-term shifts a bit the second intermediate, but does not affect the barriers. 

\subsubsection{Native strings probabilities suggest the structure of the intermediate minima.}
A complementary, two-dimensional picture of the free energy landscape can be found in Figure \ref{fig:land}, where the $\nu_{i,j}$ of Eq.~\ref{eq:nu-cappedstrings} are reported, at refolding conditions. Each point $\nu_{i,j}$ corresponds to the probability to find a native string, starting precisely at $i$ and ending at $j$. 
It is easy to identify the native spot at the bottom right corner, and the isolated short structures represented by the short strings close to the diagonal. In addition to those, five extra spots of intermediate structure can be singled out, with the three sitting at the corners being more pronounced: the central ones (roughly centered  at (32,92) and (32,106)) corresponds to the first intermediate in Fig.~\ref{fig:eq}, at $\rho \approx$  0.5 , while the others, displaced towards the N-term (the spots around (5,92) and (5,106)) or C-term (the region centered at (32,115)) respectively, are represented by the intermediate around $\rho$ = 0.75 in Fig.~\ref{fig:eq}. Interestingly, mutants involving contacts at the N-term or C-term present different probabilities at the intermediate spots, and in the regions connecting them, suggesting that also the pathways could be different between the different species. 

Notice though that correlations between isolated structured regions are neglected by construction:   the fact that, e.g., strings $(i,j)$ and $(k,l)$ (with $i<j<k<l$) appear with high probability in Fig.~\ref{fig:land}, does not imply that the configurations containing native structure at both regions $(i,j)$ and $(k,l)$ are especially likely. So, even if these two-dimensional profiles already suggest possible pathways and folding mechanisms, they do not allow a quantitative characterization of the kinetics, and a detailed study of the latter must be performed independently, as in the following section.

\begin{widetext}
$ $
\begin{figure}[!htb]
\centering
\subfigure[WT]{
\includegraphics[width=0.3\textwidth,angle=-90]{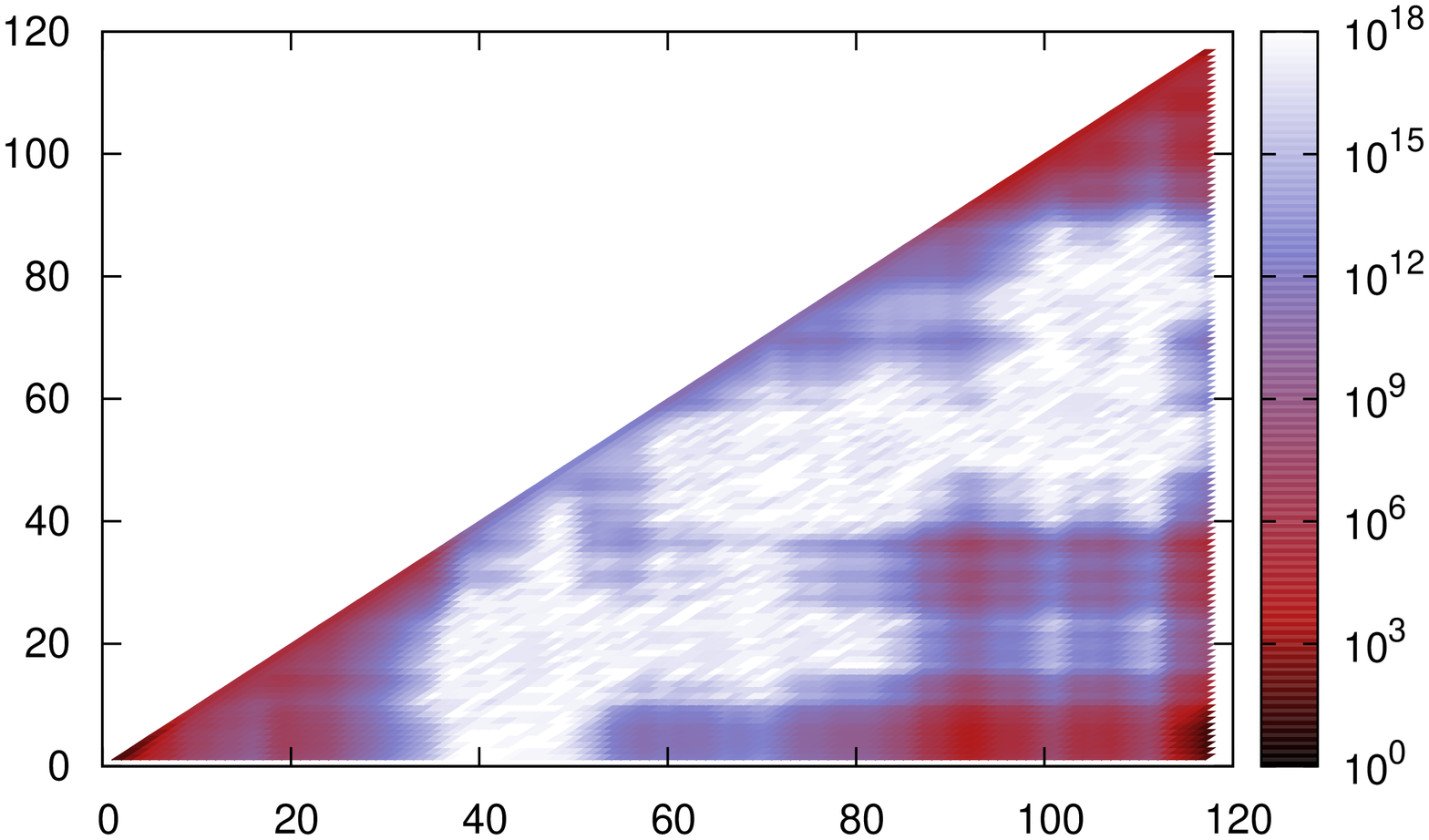}}
\subfigure[]{
\includegraphics[width=0.3\textwidth,angle=-90]{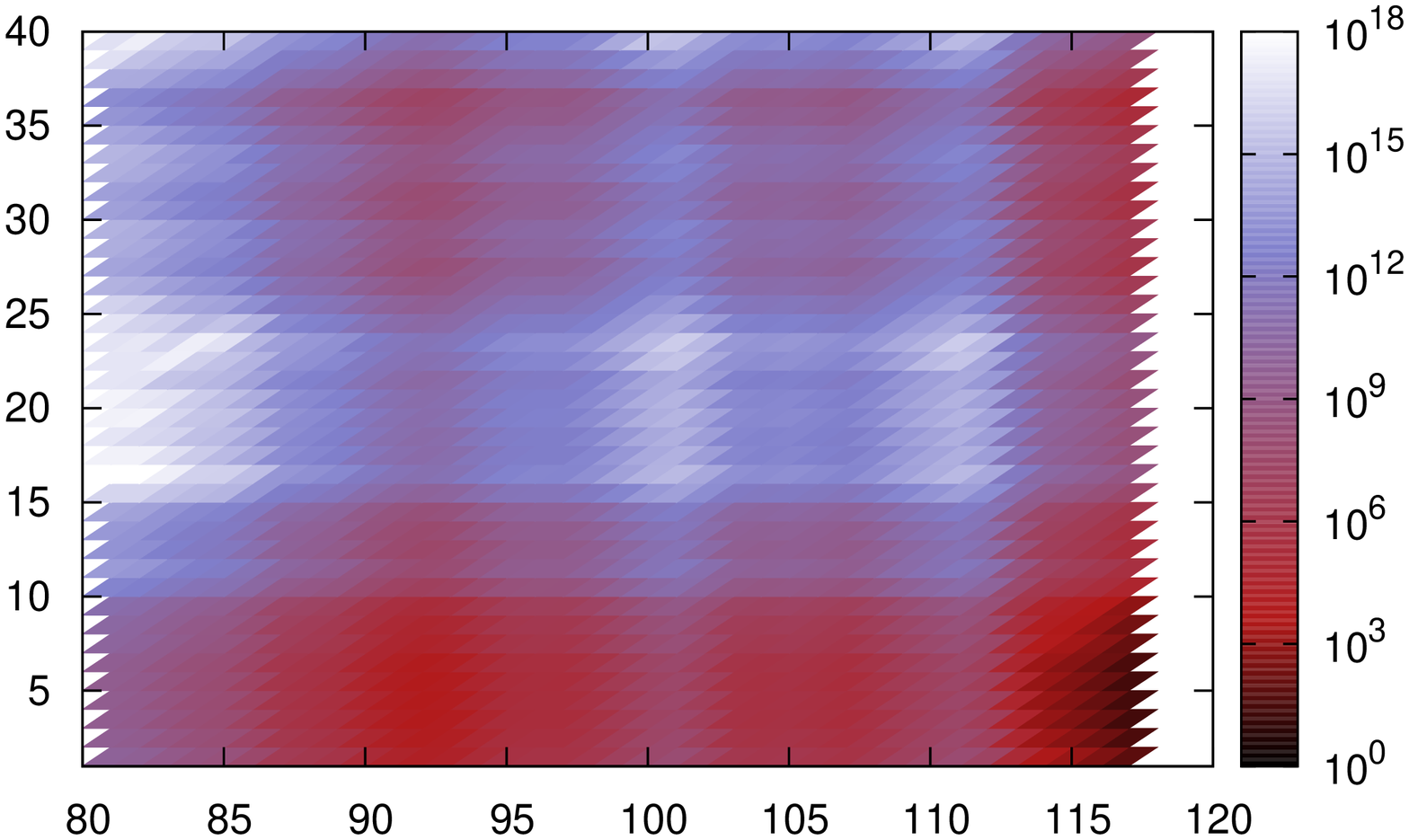}}\\
\subfigure[$S_3$]{
\includegraphics[width=0.3\textwidth,angle=-90]{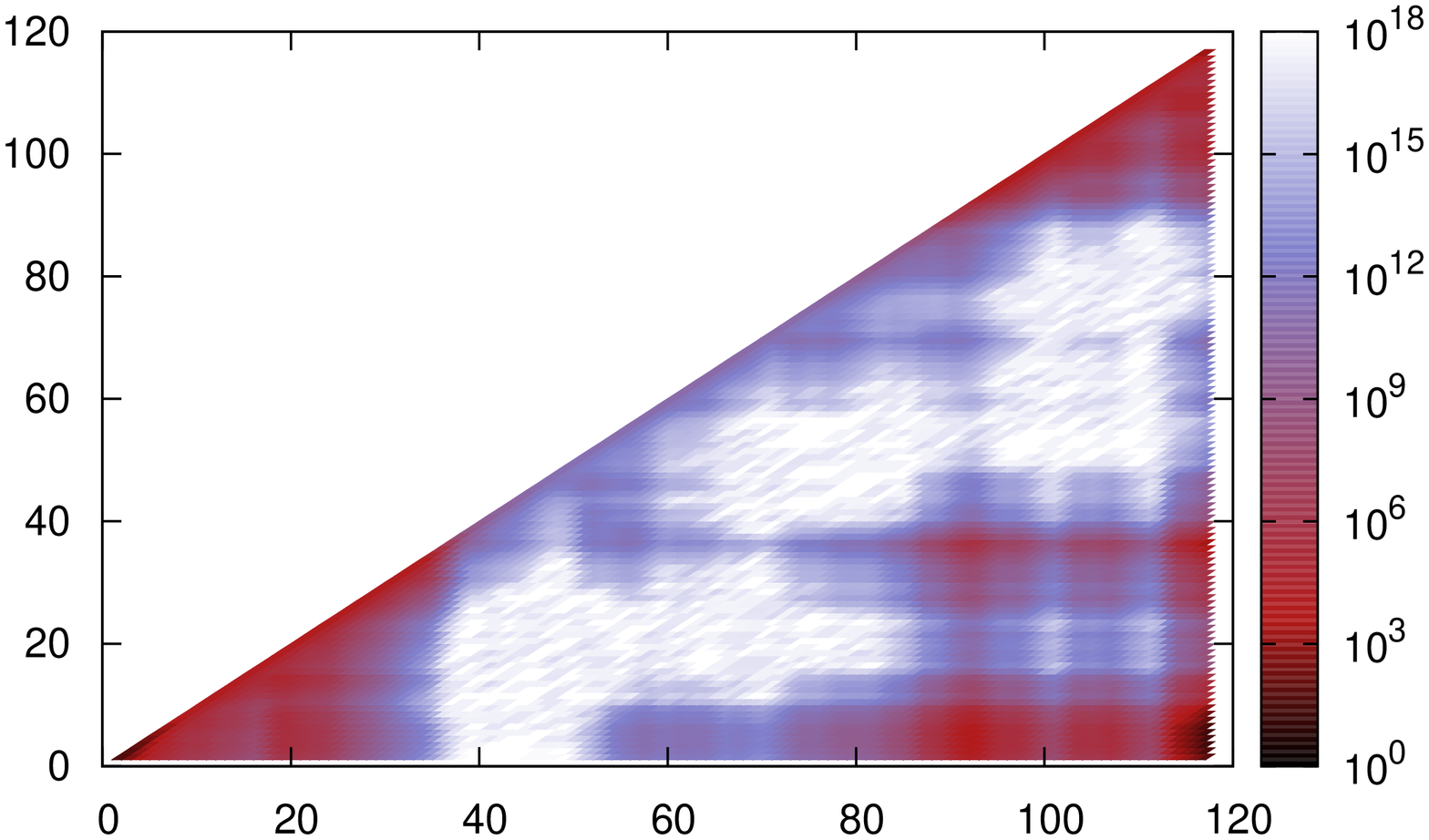}}
\subfigure[]{
\includegraphics[width=0.3\textwidth,angle=-90]{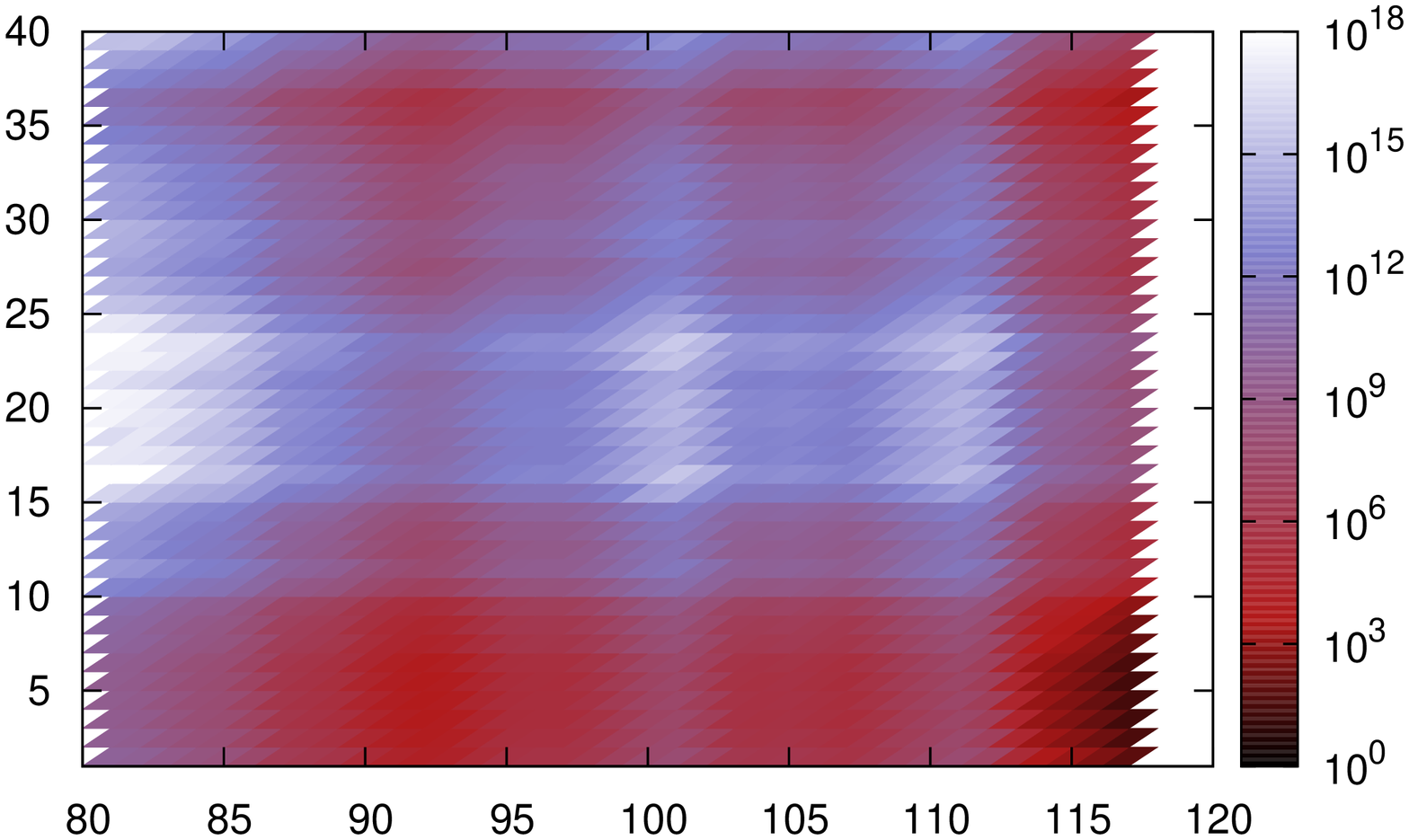}}\\
\subfigure[$S_7$]{
\includegraphics[width=0.3\textwidth,angle=-90]{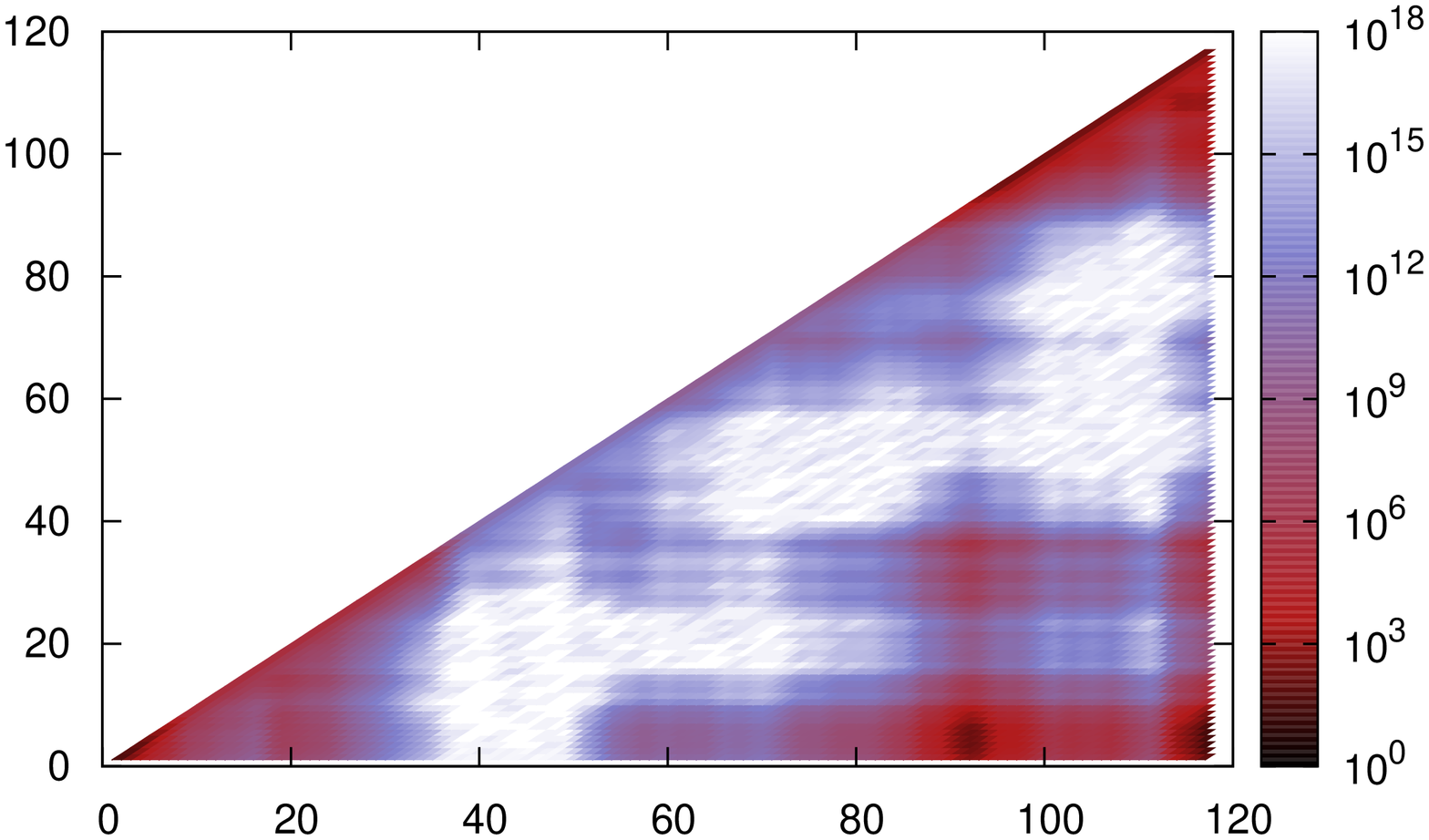}}
\subfigure[]{
\includegraphics[width=0.3\textwidth,angle=-90]{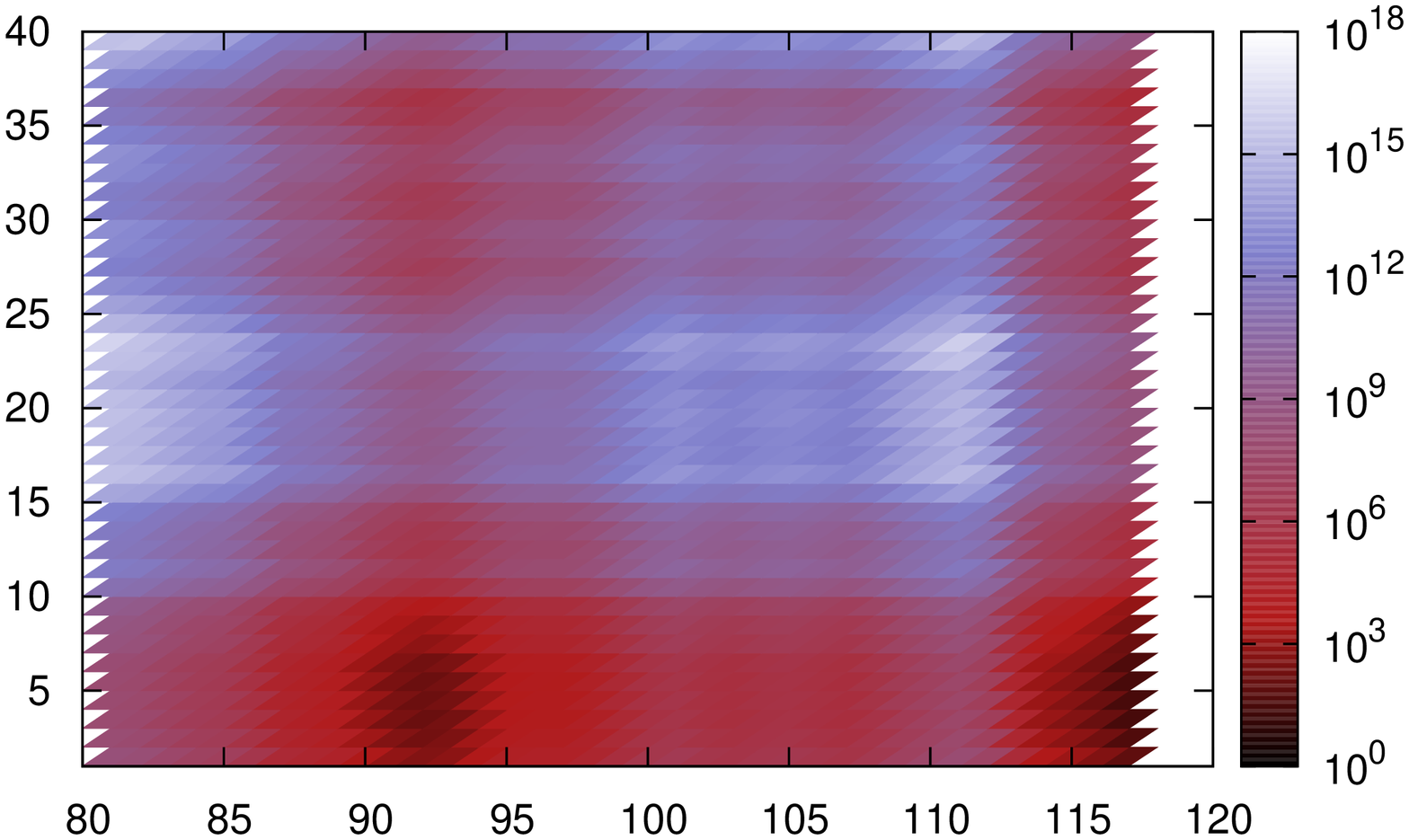}}\\
\caption{\label{fig:land} Inverse native strings probabilities $1/\nu_{i,j}$ (Eq.~{\protect \ref{eq:nu-cappedstrings}}), at folding conditions $c=0$, for the wild type and two mutants. On the right: detail of the bottom right region of the maps, close to the native structure. 
Although these 
maps  cannot be directly related to the folding kinetics, 
the qualitative folding mechanisms emerging 
from them agree with the quantitative data from the MC kinetics (see Sec.~{\protect \ref{subsec:dynamics}} below). The region around position (30,92) corresponds to the formation of a central nucleus  encompassing the two central ankyrin repeats. From there, the native state at the bottom right can be reached either elongating rightwards and then downwards ($a_f$ pathway) or downwards and then rightwards (N-term, $b_f$ pathway). 
On both pathways, after the formation of the central nucleus, there are two regions of low probability, suggesting early and late barriers on both pathways. It is possible to guess 
that the former pathway will be enhanced in the $S_3$ mutant, and the latter in $S_7$.
}
\end{figure}
\end{widetext}

\subsection{KINETICS}
\label{subsec:dynamics}
\subsubsection{Effective two-state behavior emerges despite pathway heterogeneity.}

\begin{figure}[!htb]
\centering
\subfigure[ Folding]{
\includegraphics[width=0.26\textwidth,angle=-90]{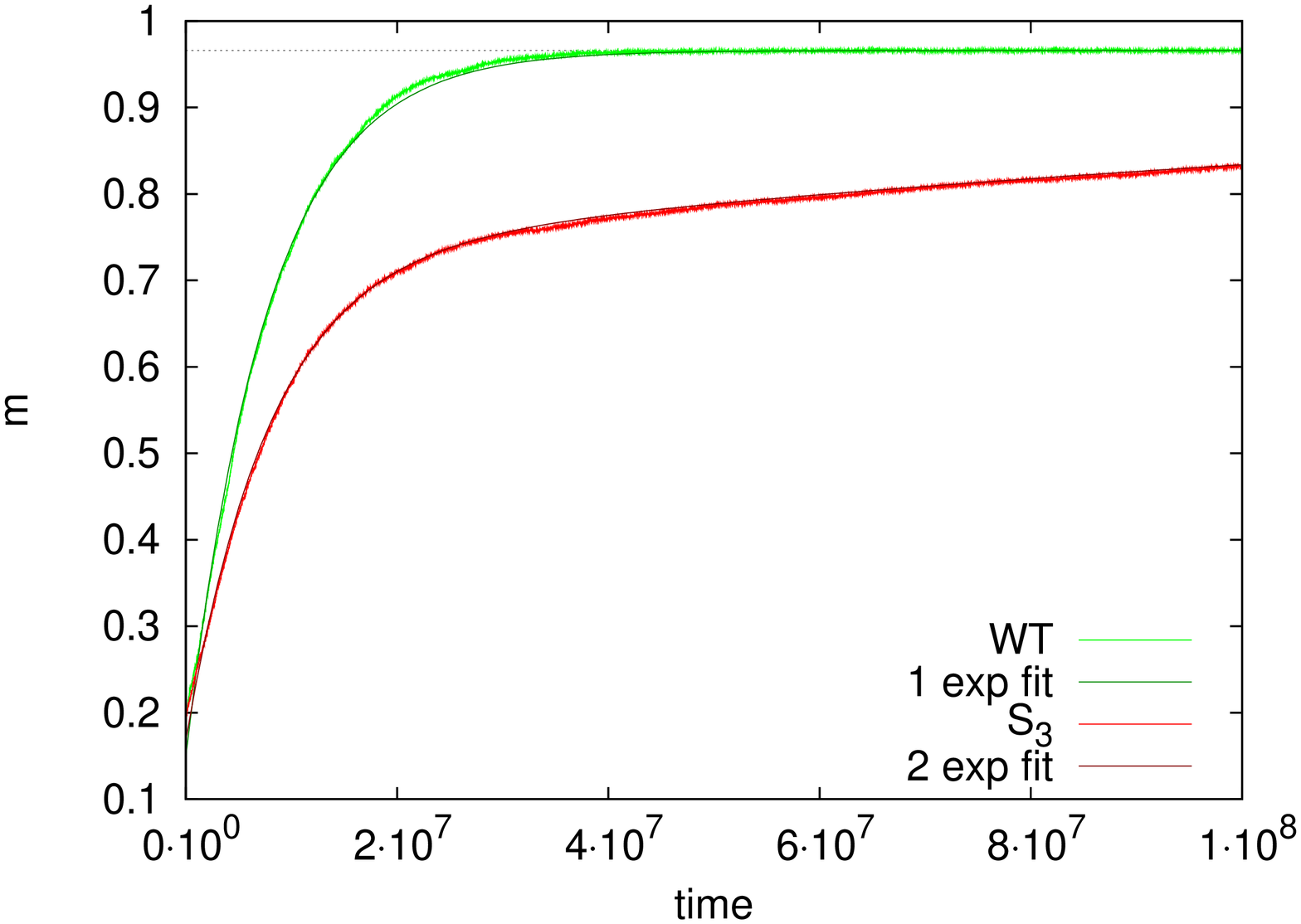}}\\
\subfigure[ Unfolding]{
\includegraphics[width=0.26\textwidth,angle=-90]{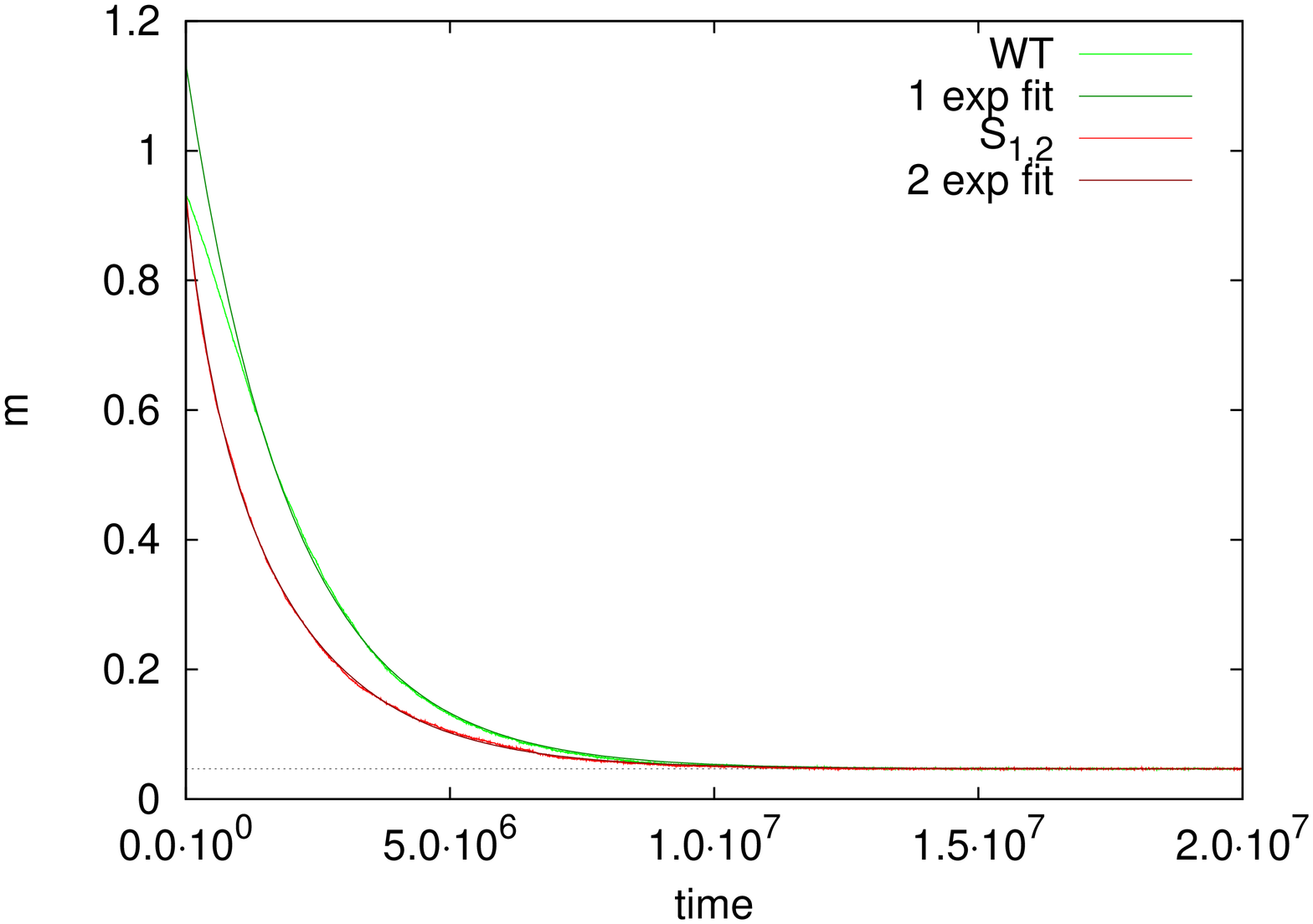}}\\
\caption{\label{fig:evo}
Plot of the fraction of native residues $m(t)$ as a function of time,  
for the folding (a) and the unfolding (b) process, for the wild-type  and for two mutants. The wild type protein shows a single-exponential behavior, also common to the majority of the mutants. The two mutants here are chosen to represent the two-exponential behavior, for comparison.  The average value was calculated over 2000 molecules simulations. 
}
\end{figure}

\begin{figure}[!htb]
\centering
\subfigure[]{
\includegraphics[width=0.32\textwidth,angle=-90]{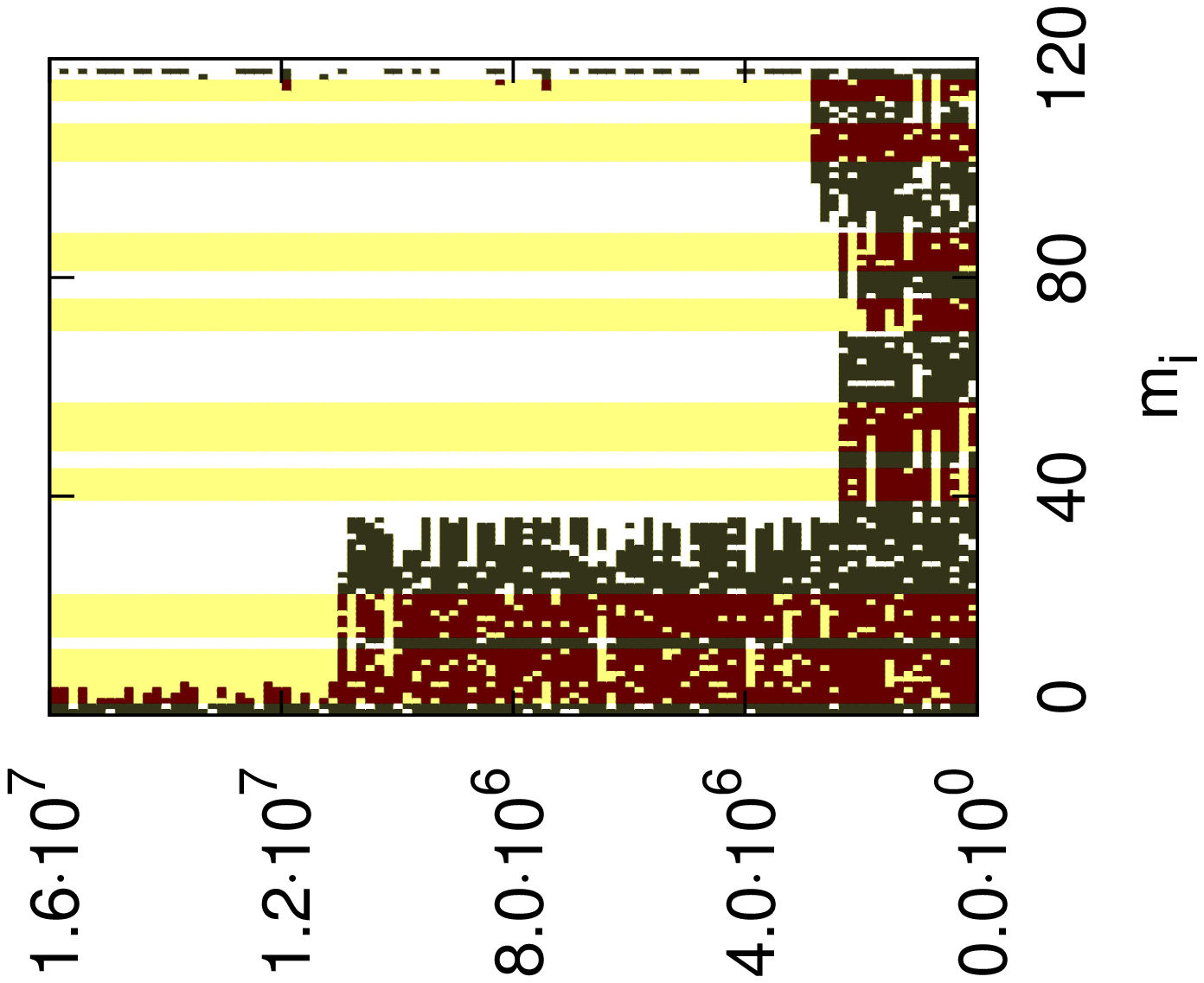
}}
\subfigure[]{\includegraphics[width=0.32\textwidth,angle=-90]{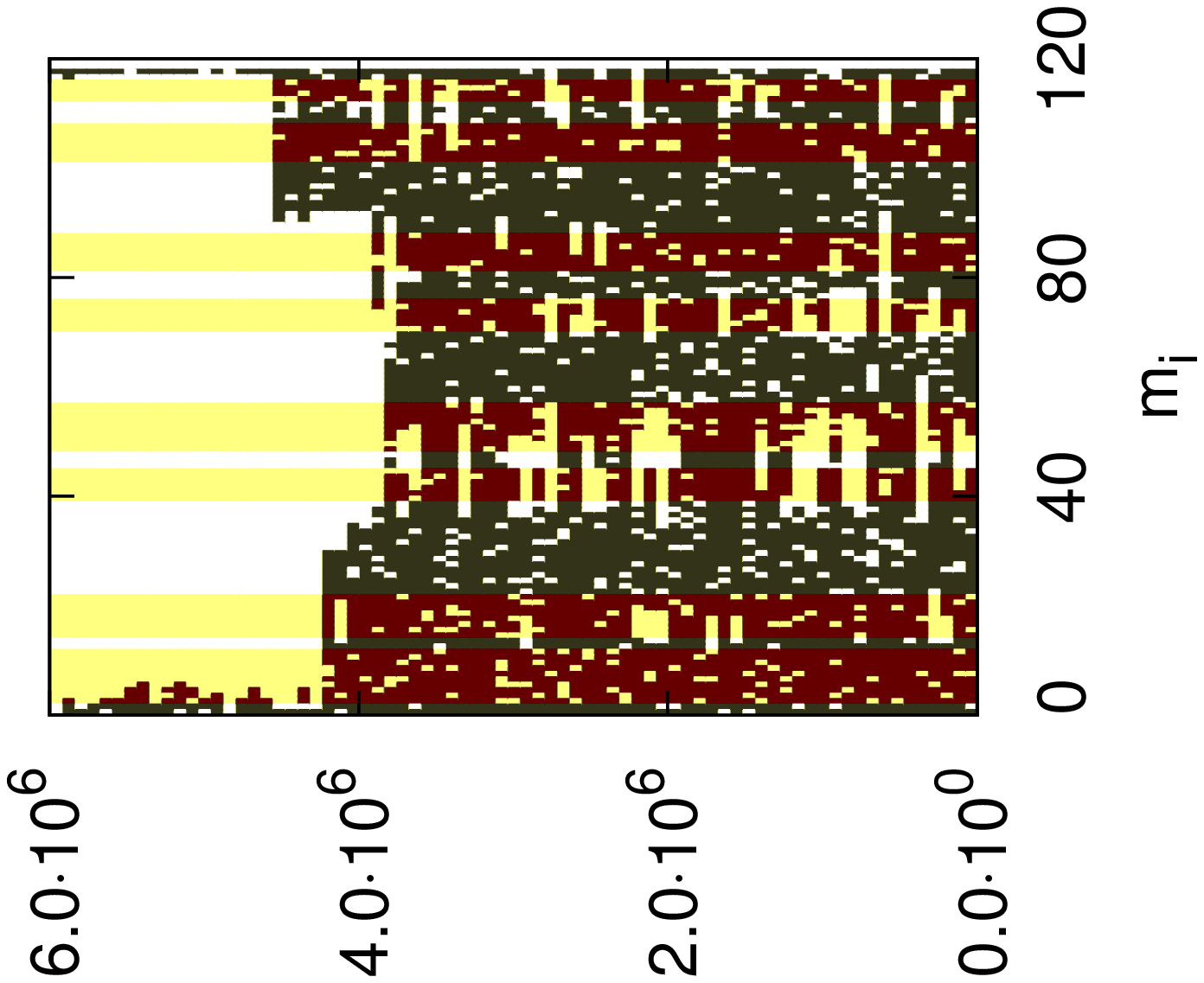
}}\\
\subfigure[]{\includegraphics[width=0.33\textwidth,angle=-90]{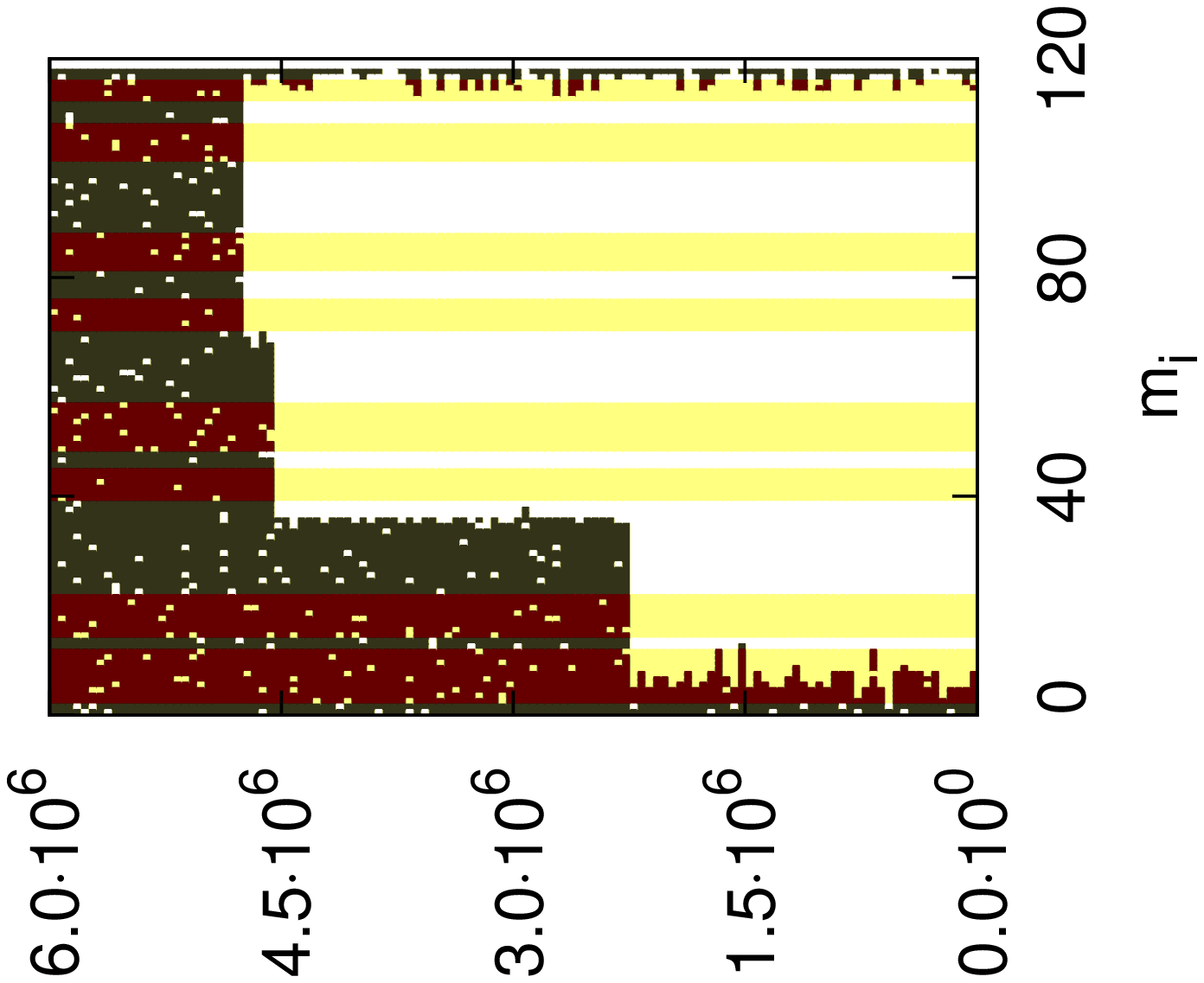
}}
\subfigure[]{\includegraphics[width=0.33\textwidth,angle=-90]{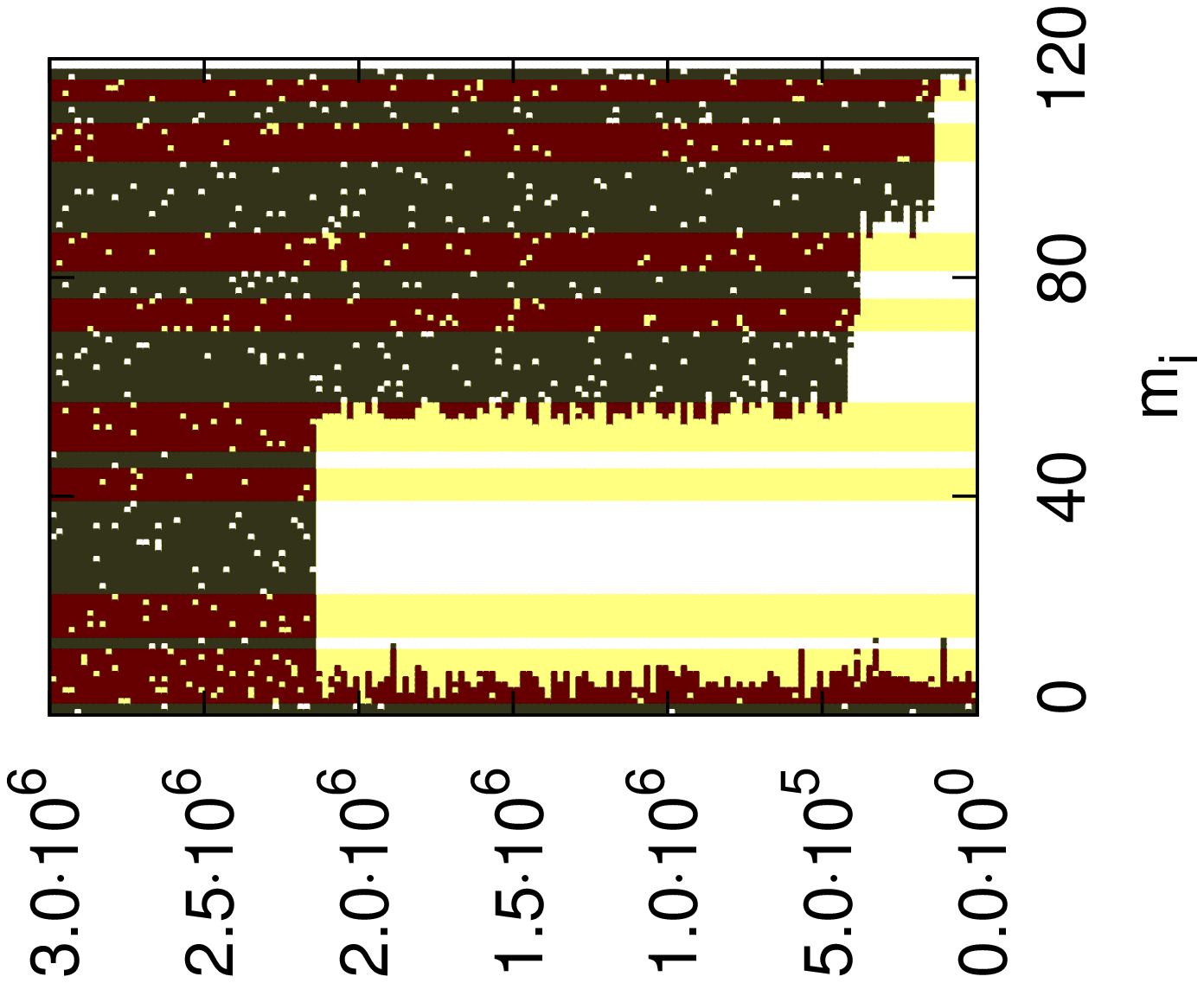}}\\
\caption{\label{fig:spaccaocchi} 
Single-molecules simulations of folding (panels a,b) and unfolding (c,d) events.
Panels a,c: relaxation along pathway ``a''; panels b,d: relaxation along pathway ``b'' (see text for details).
The residue state $m_i(t)$ is reported
with differents colors: black if unfolded ($m_i=0$) and white if native ($m_i=1$).
The vertical stripes indicate the positions of the $\alpha$-helices.
}
\end{figure}

Monte Carlo simulation of both folding and unfolding of an ensemble of 2000 molecules of the wild type species reveal a single-exponential kinetics, as can be seen in Fig.~\ref{fig:evo}, which is in agreement with the results in \cite{Lowe2007a}.
This simple behavior is apparently at odds with the multiple minima landscape reported in the free-energy profiles: to gain some detailed insight on how these characteristics can be simultaneously present, we have studied the relaxation events of individual molecules.
Some representative examples of single-molecule relaxations for the wild type species are reported in Fig.~\ref{fig:spaccaocchi}, for the folding and unfolding case. 
We have seen that, neglecting the ubiquitous structure fluctuations, it is possible to identify some precise patterns in the relaxation process: typically, the folding evolves through the stabilization of a central nucleus of four helices (second and third ankyrin repeat), after several events of formation of transient structures, involving at most individual repeats. The formation of structure at the interface between the second and third repeats typically triggers the immediate stabilization of both of them (even though this might be an artefact of the model, that just considers interactions if they take place within a native string). 
Then, the folding proceeds either in the C-term direction  or towards the N-term, and finally it reaches completion to the native state.
The unfolding at high denaturant concentration proceeds in a similar, but not perfectly symmetric way: there are still two possibilities, depending on which end unfolds first, but then the unfolding reaches completion abruptly, with an almost simultaneous unfolding of both the central part and the opposite end of the protein.
To characterize in a more quantitative way these behaviors, we have identified, for each single molecule trajectory, the rate limiting steps, by inspection of the time differences $\Delta t$ between stabilizations of different strings $t_{\alpha,\beta}^{(f)}$ along the folding or unfolding pathway, and identification of the biggest $\Delta t$ between consecutive stabilizations. We have also classified the folding (unfolding) pathways by identifying the formation (disruption) of some key strings that trigger the successive events towards the N or C term.
In both folding and unfolding case, we can distinguish two pathways, that we call $a_x$, $b_x$ where $x=f,u$ for the folding and unfolding case. Pathway $a_x$ is characterized by the C-terminal part getting structured earlier in the folding process, and disrupting later in the unfolding one.  On the contrary the $b_x$ pathway favors earlier stabilization of the N-terminal part in the folding process, and its longer persistence in the unfolding one. 
However, folding and unfolding pathway of the same kind do not coincide, so that we distinguish them with the $f,u$ labels. The detailed definition is as follows: for the folding process, we find that the event triggering the $a_f$ pathway is the formation of a native string encompassing helix 4 to 7 before  that of a native string from helix 2 to 5; the opposite order characterizes the $b_f$ pathway. 
For the unfolding process,  $a_u$ is characterized by the contacts between helices 6 and 7 lasting more than those between helix 2 and 3, while the order is reverted in $b_u$.
We have seen that with the above definitions, it is possible to classify clearly and uniquely  all the single-molecule relaxations (of the wild type and of the mutated species, see below) as belonging to either pathway.

These results are summarized in Table \ref{tab:wildtype}, where the rate and amplitude for the one-exponential fit and the fraction of molecules through the $a$ and $b$ pathways are reported.
We find a dominance of the pathway $a_f$ through the C-term over the $b_f$, and a slight dominance of  $b_u$ over $a_u$, pointing out that there may be some differences in the topography of the energy landscape in folding and unfolding conditions.  
\begin{table}[!htb]
\begin{tabular}{c|cc|cc}
\hline
\hline
& $k_1$ & $\vert c_1 \vert$ & $a$ & $b$ \\
& ($10^{-7}$)&  &(\%) & (\%) \\
\hline
folding  & 1.279(9)& 0.814(4) & 80.2 & 19.8 \\
unfolding& 4.76(8) & 0.965(4)&  40.2 & 59.8 \\
\hline
\hline
\end{tabular}
\caption{\label{tab:wildtype} 
 Rates and amplitudes (from 1-exponential fits)  and fraction of molecules that select folding
pathway $a$ and $b$, in the case of folding and unfolding relaxations. The errors have been estimated by dividing the proteins in 10 groups of 200 molecules each, evaluating a rate and amplitude from the fit of the average signal of each group, and calculating the mean and deviation of the mean of the resulting population of rates and amplitudes.}
\end{table}

%
%

How is it possible that two different pathways are present, while the folding and unfolding appear as two-state processes? The difference in the fraction of molecules following either pathway suggests that there is a little  difference in the free-energy barrier that they have to surmount. This difference cannot be huge, since in that case it would result in rates along each pathway differing by order of magnitudes, which in turn would imply fluxes by just one channel. Moreover, the fact that several minima, connected by different barriers, are found in the free-energy profiles, but the relaxation kinetics is simply exponential (two-exponential fits fail to produce reliable results due to overfitting, data not shown), implies that either the rate limiting step is represented by crossing the first barrier along the pathway, effectively masking the other jumps, or the different barriers are associated to very similar rates. 
\begin{figure}[!htb]
\centering
\subfigure[pathway $a_f$, 1604 molecules]{\includegraphics[width=0.176\textwidth,angle=-90]{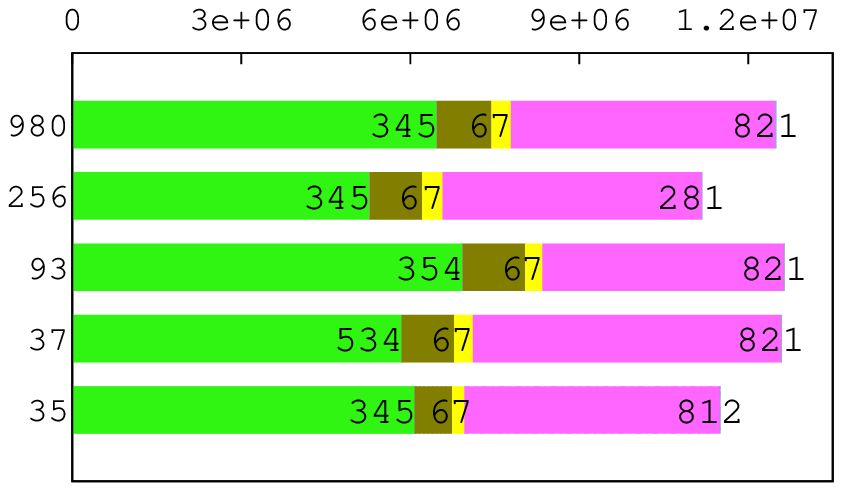}}
\subfigure[pathway $b_f$, 396 molecules]{\includegraphics[width=0.176\textwidth,angle=-90]{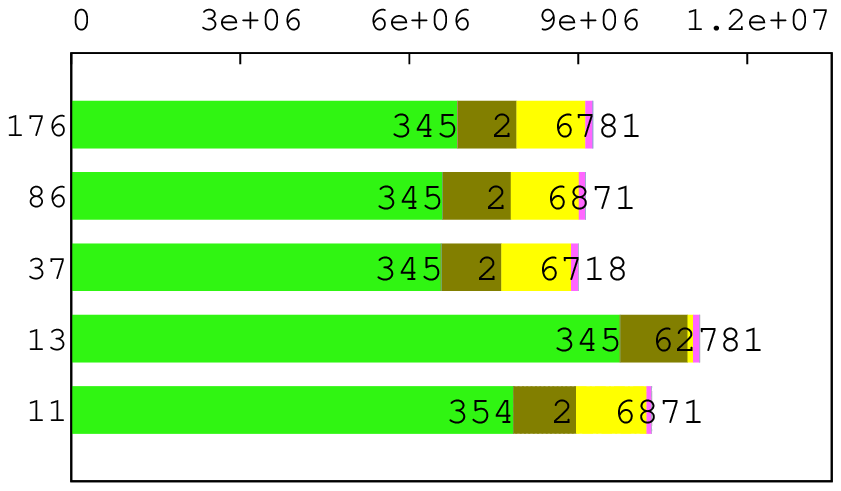}}\\ 
\subfigure[pathway $a_u$, 804 molecules ]{\includegraphics[width=0.177\textwidth,angle=-90]{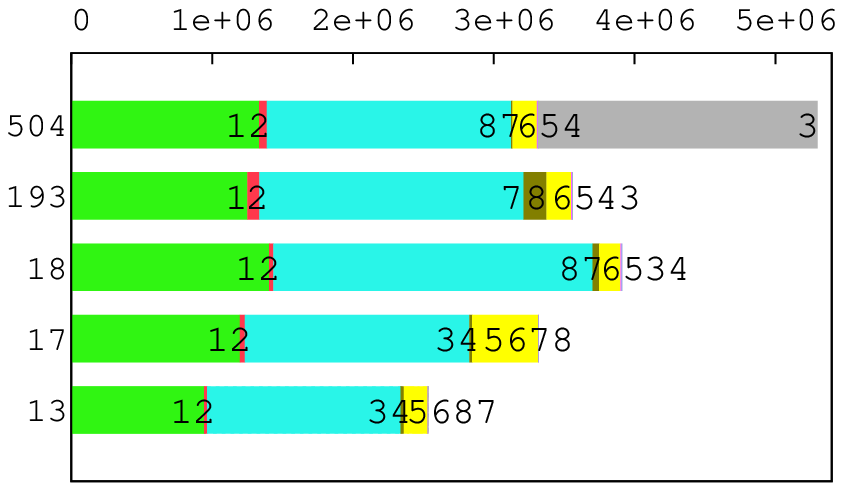}}
\subfigure[pathway $b_u$, 1196 molecules]{\includegraphics[width=0.177\textwidth,angle=-90]{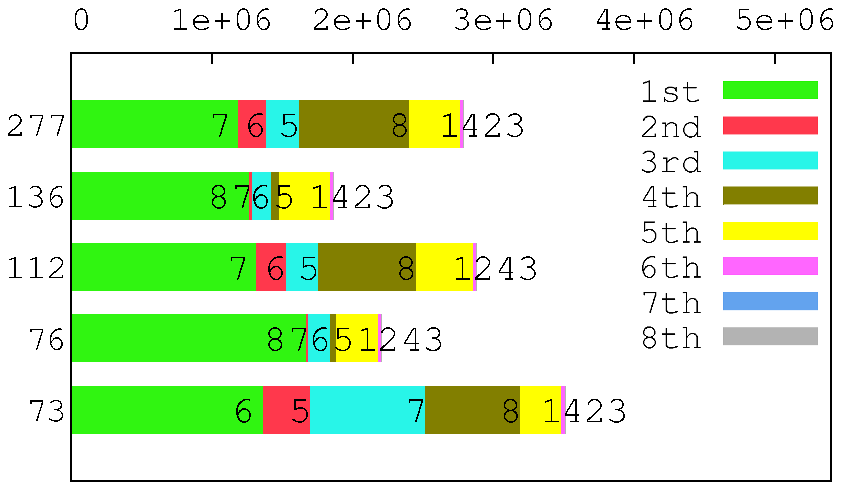}}\\
\caption{
Patterns of helices stabilization in the folding process (top) and of their disruption in the unfolding one (bottom), from the analysis of 2000 single molecule relaxations, both for folding and unfolding events. Just the five most representative patterns are reported for each folding or unfolding pathway. In the top panels, the horizontal axis represents $\langle t_{\alpha,\alpha}^{(f)} \rangle$, that is, the last time (in units of MC sweeps) that the helix $\alpha$ turns completely folded in the simulation. In the bottom panel, the corresponding quantity $\langle t_{\alpha,\alpha}^{(u)} \rangle$ for unfolding is reported. The averages are performed on all the molecules $n_s$ following the same succession of events $s$, which can be read in the labels of the bars; the number $n_s$ is reported at the left of the y-axis. The color code is the same for all panels, and refers to the order of helix stabilization (destabilization in the unfolding case). Missing colors, as well as grouping of the corresponding labels, indicate that a group of helices folds (or unfolds) almost at the same time. When this grouping takes places, it is quite common to  find patterns that differ just for the permutation of grouped labels: e.g. this is the case for the $a_f$ pathways reported in panel (a), where all the patterns are small variation of the same scheme in four steps: folding of helices 3, 4, 5 basically at the same time, then stabilization of helix 6 and then 7 after a short time, and finally completion of the folding. 
}
\label{fig:avetimes}
\end{figure}
This picture is confirmed by the analysis of the average times of helix stabilization (or destabilization, in the unfolding process), reported in Fig.~\ref{fig:avetimes}, where the most representative patterns of secondary structure formation are reported separately  for both the folding and unfolding pathways. It is clear from the top panels that the folding pathways are characterized  by  the formation of helices 3, 4, 5 basically altogether, around t=$6\cdot 10^6$, followed by the extension, in another million of time steps, toward helices 6 and 7 in pathway $a_f$, or helix 2 in pathway $b_f$. Then, the rest of the structure folds almost at once. In the folding process, the longest time is associated to the formation of the initial nucleus, and the second longest one (and close to the former), to the completion of  the folding along pathway $a_f$.

The unfolding process presents as well some common schemes: in the dominant $b_u$ pathway, unfolding proceeds from the C-term (helices 7, 6 and 5), passes through the last and first helix, and finally affects the rest of the N-terminal part. The $a_u$ pathway presents more variability, but the dominant mechanism is given by the $a_f$ pathway covered in the opposite direction, and ending with the central group of helices 3, 4 and 5. Notice that in both the unfolding pathways, the second repeat appears as the last to unfold, and  the longest time is usually associated to the unfolding of the last repeat, and in particular of helix 7.

Fig.~\ref{fig:avetimes} suggests a clear picture of how the folding and unfolding proceed along the $a_x$ and $b_x$ pathways
and gives an idea of what are the rate limiting steps, even if it does not inform on the detailed structure of the transition states and nuclei, because the latter could contain partially structure helices and hence need not coincide with a collection of fully formed helices, while the times $t_{\alpha,\alpha}^{(f)}$, $t_{\alpha,\alpha}^{(u)}$  inform on when the helix $\alpha$ becomes stably structured or unstructured as a whole, and are affected by structural fluctuations within the helix.
Moreover, averaging the times $t_{\alpha,\alpha}^{(f,u)}$ (that may vary a lot from molecule to molecule) gives no information about the time evolution of any observable.

For these reasons, the picture coming from Fig.~\ref{fig:avetimes} must be checked and confirmed by further analysis. To this end,  
 we have simulated the effect of  mutations, as explained in Section \ref{sec:methods}, and performed the same analysis as for the wild type species.

\subsubsection{Analysis of simulated mutants confirms the proposed kinetics mechanism, with pathway heterogeneity.}

Table \ref{tab:pert} summarizes the results for the kinetics of the mutants: most of the times, the $m(t)$ signal can adequately be fitted with just one exponential (with rate $k_1$) both in the folding and unfolding case, while some mutants present a second, faster  phase $k_2$, especially in the unfolding case.

%

\begin{table}[!htb]
\begin{tabular}{c|cccc|cccc}
\hline\hline
&\multicolumn{4}{c|}{folding}&\multicolumn{4}{c}{unfolding}\\
\hline
& $k_1$ & $k_2$ & $a_f$ & $b_f$& $k_1$ & $k_2$ & $a_u$ & $b_u$\\
& ($10^{-7}$) &($10^{-7}$)& (\%) & (\%) &($10^{-7}$) &($10^{-7}$)& (\%) & (\%)\\
\hline
WT		&1.29	& -	&80.2 	&19.8	&5.08	& -	&40.2 &59.8	\\
$S_{1,2}$	&1.22	& -	&85.1	&14.9	&5.20	&32.0	&85.7 &14.3	\\
$S_3$   	&0.058	&1.26	&98.7	&1.3	&5.00	& -	&39.6 &60.4	\\
$S_{3,4}$	&0.116	& -	&85.9	&14.1	&6.74	& -	&38.8 &61.2	\\
$S_{5,6}$	&1.22	& -	&80.2	&19.8	&4.92	& -	&41.0 &59.0	\\
$S_7$		&1.18	& -	&45.3	&54.7	&6.45	& -	&29.9 &70.1	\\
$S_{7,8}$	&1.26	& -	&61.7	&38.3	&6.95	&47.8	&1.7  &98.3	\\
$S_{1,4}$	&1.28	& -	&80.5	&19.5	&5.02	&21.4	&88.2 &11.8	\\
$S_{3,6}$	&1.25	& -	&78.2	&21.8	&5.13	& -	&39.1 &60.9	\\
$S_{5,8}$	&1.34	& -	&69.1	&30.9	&6.99	&50.9	&1.6  &98.4	\\
\hline\hline
\end{tabular}
\caption{\label{tab:pert} Rate and fraction of molecules that
select folding
pathway $a$ and $b$ in the case of folding and unfolding dinamics for differt
types of perturbation, as compared to the wild type (WT). Here the rates are calculated by fits on the whole ensemble of 2000 proteins. Missing entries in column $k_2$ mean that the 1-exponential fit was sufficient, and the 2-exponential one  would produce overfitting.}
\end{table}

\paragraph{Folding kinetics.} 
We see that in the folding case, mutations affecting helices 2,3 and 7,8 do not change the rate much, but they are those that tune the flow along the two pathways. Mutations at the C-term cause the biggest pathway shifts, in agreement with the experimental results  \cite{Lowe2007}.
Mutations in the central nucleus (e.g.  $S_{3,4}$) affect the rate , but do not change much the distribution along the two pathways. 
Mutants $S_{5,6}$, $S_{3,6}$ 
are associated to contacts that do not stay at  a barrier top: their destabilization weakly affects the rates and pathways. 

The above results confirm that 
the formation of the central nucleus
of three helices 3,4 and 5 is the rate limiting step for the folding process, followed by a growth of the nucleus towards
the C or N term (pathways $a_f$ and $b_f$ respectively). 
The barriers associated to these pathways are smaller enough than the nucleation one, so that most of the perturbations that shift the flow cannot ``promote'' these barriers to be the highest one. As a result,  the above sequence of events is preserved in the mutants, and the folding rates are little affected and quite similar to the WT one, while the distribution of the flow changes according to which pathway is destabilized. The behavior of $S_{3,4}$ is consistent with the above observations: here the central nucleus is destabilized, which results in a slower rate with a moderate change in the flows. The only exception to this picture is mutation $S_3$, that affects all the contacts (local and nonlocal) of residue 32, located in the loop between the first and second repeat. The destabilization of these contacts  leaves the rate for the formation of the central structure unchanged, but produces a second, slower rate (that therefore we shall interpret as the folding rate), corresponding to the last steps of the  folding along pathway $a_f$, of stabilization of the structure at the first repeat. Accordingly, the corresponding folding flow goes almost completely through the $a_f$ pathway.
Consistent with this intepretation, $S_{1,4}$ does not involve relevant changes in either the folding rate or fluxes, and  appears downhill with respect to the crossing of the last barrier along the $a_f$ pathway.

\paragraph{Unfolding kinetics.}

As in the folding, mutations affecting the external regions (first and last ankyrin repeats) cause the biggest changes in the flow through the pathways; these changes agree with those observed in the folding case: a mutation causing a larger flux towards the $a_f$ pathway in folding will also cause an increase in the $a_u$ flux in unfolding, though of different magnitude. Moreover, these mutations  are accompanied by the appearence of a second, faster rate, signaling that a part of the structure unfolds before the rate limiting step; the latter rapidly leads  to the completion of the unfolding. 
On the other hand, mutations affecting the central region do not cause major changes in either the rate or the flux distribution along the two pathways, with respect to the wild type.
Interestingly, in the $S_3$, $S_{3,4}$ and $S_{5,6}$ species the changes in the flux have opposite sign in the folding and unfolding processes, again suggesting that the choice of the pathway is not controlled by the central repeats.  
As Table \ref{tab:pert} suggests, the presence of two rates, in either folding  or unfolding, is not related to the two different pathways for the kinetics, as one could naively think at the beginning. This is even clearer in Table \ref{tab:pathwayrates}, where the single or two-exponential fits are performed separately on the subsets of molecules following either pathways: in general, the need for a  two-exponential fit for the full ensemble is associated to the presence of two rates in either the $a$  or  the $b$ pathway. 


\begin{table}[!htb]
\begin{tabular}{l|cc|cc|c|cc}
\hline\hline
&\multicolumn{2}{c|}{$a_f$} & \multicolumn{2}{c|}{$b_f$} &$a_u$ & \multicolumn{2}{c}{$b_u$}\\
\hline
& $k_1$ & $k_2$ & $k_1$ & $k_2$ & $k_1$ & $k_1$ & $k_2$\\
& ($10^{-7}$) & ($10^{-7}$) & ($10^{-7}$) & ($10^{-7}$) & ($10^{-7}$)& ($10^{-7}$)& ($10^{-7}$)\\
\hline
WT       &  1.27 & -     & 1.36 & -     & 4.53 & 5.93 & -    \\
$S_{1,2}$&  1.19 &  -    & 1.40 & -     & 5.14 & 27.8 & -    \\
$S_{3}$  &  0.056& 1.24 & 0.53 & 3.68 & 4.10 & 5.76 & -    \\
$S_{3,4}$& 0.115& -   & 0.109 & -    & 6.74 & 6.73 & -    \\
$S_{5,6}$& 1.18 & -     & 1.34 & -     & 3.87 & 5.38 & -    \\
$S_{7}$   & 1.13 & -     & 1.18 & -     & 7.11 & 6.40 & -    \\
$S_{7,8}$&  1.25 & -     & 1.24 &  -    & 37.2 & 6.86 & 48.2\\
$S_{1,4}$&  1.25 & -     & 1.42 &  -    & 5.06 & 23.2 &  -   \\
$S_{3,6}$& 1.21 &  -    & 1.37 &  -    & 4.51 & 5.73 &  -   \\
$S_{5,8}$ & 1.30 &  -    & 1.35 & -     & 49.2 & 7.12 & 54.2\\
\hline\hline
\end{tabular}
\caption{Rates calculated on the subsets of molecules following the different pathways of folding and unfolding, as found in Table {\protect \ref{tab:pert}}.}
\label{tab:pathwayrates}
\end{table}


Another interesting thing is that the rates along the less populated pathways are usually comparable to or greater than those along the corresponding dominant one. This apparent  contradiction can be explained if one considers that, due to the restriction of the fit to a specific subset of the molecules, the resulting rate is not related to the equilibrium distribution, and it gives no information on the height of the free-energy barrier along the pathway. This can be easily understood by considering, for example, wild type molecules: after the formation of the nucleus of the two central repeats, they have to choose whether to follow the $a_f$ pathway, with a low barrier, or the $b_f$ one, with a higher one. Table \ref{tab:pert} shows that the majority of them will follow the former. Therefore, the fraction of molecules along the $b_f$ pathway will be given by those that choose that pathway early, i.e. in times shorter than the typical folding time along $a_f$, producing an apparently faster rate.  This can also be seen in the top panels of Fig.~\ref{fig:avetimes}: the formation of helix 2 in the $b_f$ pathway takes place in a time of the same magnitude as the selection of helix 6 and 7 in the $a_f$ one. After the formation of helix 2, the folding along $b_f$ is faster than the competing one. Thus, a fit restricted to the $b_f$ ensemble yields naturally a faster rate than the one found for the $a_f$ pathway. 
Indeed, we have checked that if the $b_f$ pathway is imposed  to the wild type protein, by associating an energy penalty to the formation of long native string towards the C-term, so that the protein cannot ``escape'' through the $a_f$ pathway, the resulting rate $k_{b_f}=1.20 \cdot 10^{-7}$ is lower than that observed in free wild type.



%
%
%

\paragraph{Multiple mutations}

In Ref.~\cite{Lowe2007}, Lowe and Itzhaki induce switches between the pathways by engineering multiple mutants.  In order to compare our model to those results, we have simulated the effect of those mutations applying a stabilizing or destabilizing perturbation, as explained in Section~\ref{sec:methods}.
%

The model predictions for such mutants, reported in Tables S1 and S2 of Ref.~\cite{suppmat2}, 
 show that the redistribution of the flux along the pathway upon combination of point mutations is qualitatively as expected from the discussion in the previous section, so that a mutation favouring  pathway $a$ will balance the effects of a mutation favouring pathway $b$, recovering at least partially the WT flows, even if with smaller folding and bigger unfolding rates. 
It must be noticed, though, that the ratio between the corresponding rates do not reflect the experimental results, and the model fails to give quantitative predictions of the rates.


\section{DISCUSSION AND CONCLUSIONS}

The results reported show that the WSME model reproduces qualitatively the experimental behavior: indeed, we find sigmoidal, apparently two-state-like equilibrium signals, a two-state-like kinetics with transient intermediates, and also two pathways in kinetics, characterized by the order of structure formation at the C- and the N-term. Moreover, we find that precisely targeted perturbations of the contact interactions at  different positions along the chain allow to induce pathway switches, as seen in experiments, or the stabilization of some transient intermediate resulting in a faster phase, and provide a way to probe the folding mechanisms to a great detail. 
The intrinsic complexity of the free-energy landscape of the protein myotrophin is evident already from the observation that the sigmoidal signals actually hides a three-state equilibrium (see Ref.~\cite{suppmat1}
), and from the analysis of the free-energy profiles, Fig.~\ref{fig:eq}. For this protein the latter, 1-dimensional, projection is not sufficient to suggest the details of the kinetics, since the two different pathways cannot be distinguished just on the basis of the number of native residues, which  is the natural reaction coordinate of the model. 
It is also important to notice that the free-energy profiles, in strongly denaturing or renaturing conditions, present small barriers (less than 3 $RT$), which seems at odds with the much slower rates found in the simulations. We see three possible reasons for such difference: first, in the presence of two different pathways in the configuration space, but with barriers located at similar values of the reaction coordinate and thus roughly overlapping in the projection (see Fig.~\ref{fig:land} and the relative discussion), the profile will be  always more representative of the lower of the two overlapping barriers, since, by construction, it is more representative of the states with higher Boltzmann weight. However, if, as in this case, the lower early barrier is on the $a$ pathway and the lower late barrier is on the  $b$ pathway, the true barrier on each pathway will be higher than it may be inferred from the profiles. Second, the presence of wide and almost flat regions between each minima and the following barrier slows down the rate, according to the role of the curvature of the profile in Kramers' theory. Third, the projection collects together, at the same values of the reaction coordinate, configurations that can be highly different (in terms of the Hamming distance), especially in the unfolded region. This is irrelevant to kinetics as long as the motions in the transverse directions are fast enough to be relaxed at equilibrium when considering displacements along the reaction coordinate, but this is not necessarily granted for proteins with pathway heterogeneity. Independently from which of the above possibilities is more relevant, the important message from the above observations is that any quantitative conclusion about kinetics, derived from the analysis of the free-energy profiles, must be drawn with care, especially for proteins with a pathway heterogeneity.
Yet, some hints for such a read-out of the kinetics come from the analysis of the equilibrium probability for the formation of native strings, Fig.~\ref{fig:land}: even if such probabilities do not constitute a free-energy map, and cannot be used to predict the kinetics by transition state theory or by solving a diffusion equation, they suggest the important spots playing a key role in the intermediate states, the possible folding pathways, and the way mutations can affect the kinetics, redirecting the folding and unfolding fluxes.   

The picture that emerges, and that could probably be generalized to other repeat proteins, is that in such proteins multi-minima free-energy profiles are the rule, with intermediate states related to the completion of the folding of whole repeats or substructures of them. In this framework, the cooperativity in the equilibrium unfolding reported in experiments would be attained by a ``designed'' free-energy landscape, such that in all conditions the intermediates are associated to free-energies substantially higher than those of the native and unfolded states. Such design would involve sequence-heterogeneity between the different repeats, to ensure different degrees of stability to different partially folded structures. Mutations can alter this situation\cite{Kloss2008}, and indeed we see that cooperativity is reduced by perturbing the N-term, and especially the C-term of Myotrophin. 
Two-state kinetics would most likely emerge, in such a multi-minima landscape, when the rate-limiting step coincides with the crossing of the first barrier encountered in the folding or unfolding process, and masks the crossing of the following barriers. Again, mutations may ``promote'' different barriers to the status of rate-limiting step, thus involving multi-state kinetics and/or pathway heterogeneity.

Unfortunately, the predictions of the model, especially for the kinetics, cannot be made quantitative, at least at this level of simplicity. In the model we adjust only two parameters to reproduce the temperature and concentration of the folding midpoint, and at this level of simplification, we cannot reproduce the ratio of the rates between the mutants studied in the experiments. Also, the central nucleus that we find in the folding process is not reported in the intepretation of the experimental data by Lowe and Itzhaki (that propose a three-state dominant pathway plus a two-state secondary one to interpret their data, and assume that the total relaxation rate is just the sum of the rate along the two pathways). The enhancement of the role of such a common central nucleus, that delays the choice of either folding pathway to higher values of the reaction coordinate, might therefore be a model artifact, due to the model feature of considering the interaction energies as just proportional to the number of contacts of each residue: this may effectively penalize the terminal helices, that make fewer contacts.

However, it is important to stress that the model gives predictions which go beyond the experimental results, as for instance the detailed information about the pathways, providing useful conceptual frameworks for the intepretation of the experimental data. 
An important suggestion coming from the model predictions is that folding and unfolding pathways are not necessarily the same pathway, covered in opposite directions: the different denaturant concentrations (or temperature conditions) may involve subtle but important changes in the energy landscape, so that the overall mechanisms (for instance, the two-state, two-pathway kinetics)  do not change, but the details of the pathways do. Indeed, in the presence of two possible pathways and a strong bias towards folding (or unfolding), it is most likely that at any ``fork'' in the pathway the protein will follow the trail with the smaller barrier at that point, and will be stuck on it, since backwards jumps will be strongly suppressed. In the present case, the lower folding barrier at lower values of the reaction coordinates on the $a_f$ pathway, and the lower unfolding barrier at higher values of the reaction coordinates on the $b_u$ pathway, produce heterogeneity in the folding and unfolding pathways. Independently on how realistically this mechanism describes the behavior of myotrophin, it
is an important warning for the design of simple interpretation frameworks for experimental results: if on the one hand, a phenomenological model must be kept as  simple as  possible, to avoid overfitting and the introduction of too many parameters, on the other hand, the use of simple models as the present one, with very few free parameters, may represent a useful alternative to get a grasp on the key mechanisms of the folding and unfolding process.

\begin{acknowledgments}
P.~B. and M.~F. acknowledge support from the Spanish Ministry of Cience (MICINN) through grant FIS2009-13364-C02-01. M.~F. holds a fellowship by the Diputaci\'on General de Arag\'on (B045/2007).
The numerical calculations were run with in-house software on the BIFI computer cluster. 
\end{acknowledgments}

%


\end{document}